\documentclass[a4paper, fleqn, usenatbib]{mnras}
\usepackage{newtxtext,newtxmath}
\usepackage[T1]{fontenc}
\usepackage{ae,aecompl}

\usepackage{graphicx}	% Including figure files
\usepackage{amsmath}	% Advanced maths commands
\usepackage{hyperref}
\graphicspath{{./fig/}} % define figure path
\usepackage[normalem]{ulem}

\newcommand{\HII}{\rm H~{\sc ii }}
\newcommand{\ctworay}{\texttt{C$^2$RAY}~}
\newcommand{\mpc}{ Mpc}
\newcommand{\secref}[1]{\S~\ref{#1}}

\title[21cm Segmentation U-Net]{Deep learning approach for identification of {\HII} regions during reionization in 21-cm observations}
\author[M. Bianco et al.]{Michele Bianco,$^{1,2}$\thanks{Contact e-mail: M.Bianco@sussex.ac.uk} Sambit. K. Giri,$^{3,2}$
Ilian T. Iliev$^{1}$, Garrelt Mellema$^{2}$ \\
$^{1}$Astronomy Centre, Department of Physics \& Astronomy, Pevensey III Building, University of Sussex, Falmer, Brighton, BN1 9QH, United Kingdom\\
$^{2}$ The Oskar Klein Centre, Department of Astronomy, Stockholm University, AlbaNova, SE-10691 Stockholm, Sweden\\
$^{3}$ Institute for Computational Science, University of Zurich, Winterthurerstrasse 190, 8057 Zurich, Switzerland
\\}

\date{Accepted 2021 May 21. Received 2021 May 18; in original form 2021 February 12}

\pubyear{2021}

\begin{document}
\label{firstpage}
\pagerange{\pageref{firstpage}--\pageref{lastpage}}
\maketitle
\begin{abstract}
The upcoming Square Kilometre Array (SKA-Low) will map the distribution of neutral hydrogen during reionization and produce a tremendous amount of 3D tomographic data. These images cubes will be subject to instrumental limitations, such as noise and limited resolution. Here we present \texttt{SegU-Net}, a stable and reliable method for identifying neutral and ionized regions in these images. \texttt{SegU-Net} is a U-Net architecture based convolutional neural network (CNN) for image segmentation. It is capable of segmenting our image data into meaningful features (ionized and neutral regions) with greater accuracy compared to previous methods. We can estimate the ionization history from our mock observation of SKA with an observation time of 1000 h with more than 87 per cent accuracy. We also show that \texttt{SegU-Net} can be used to recover the size distributions and Betti numbers, with a relative difference of only a few per cent from the values derived from the original smoothed and then binarised neutral fraction field. These summary statistics characterise the non-Gaussian nature of the reionization process.

\end{abstract}

%%%%%%%%%%%%%%%%%%%%%%%%%%%%%%%%%%%%%%%%%%%%%%%%%%

%%%%%%%%%%%%%%%%% BODY OF PAPER %%%%%%%%%%%%%%%%%%
\section{Introduction}
The Epoch of reionization (EoR) is a period of great importance in the study of structure formation and evolution in the Universe. During this period, the predominately cold and neutral intergalactic medium (IGM) transitioned to a hot and ionized state due to the appearance of the first luminous cosmic sources. These sources, which may have been star-forming galaxies and quasi-stellar objects (QSOs), produced the ionizing photons, which over a period of approximately 500 million years completed the reionization of the Universe \citep{Furlanetto2006CosmologyUniverse, Zaroubi2012, Ferrara2014}. 

This period is one of the least understood epochs in the history of the Universe, due to the lack of direct observations. Indirect constraints have been put on the reionization process based on observations of the Lyman-$\alpha$ forest \citep[e.g.][]{Fan2006, McGreer2011, McGreer2014}, the number density of Lyman-$\alpha$ emitters \citep[e.g.][]{Ota2008, Ouchi2010, Robertson2015}, %Lyman-$\alpha$ damping wing in 
high-z quasar spectra \citep[e.g.][]{Schroeder2013, Totani2016, Davies2018, Greig2019} and the measurement of the Thomson scattering optical depth towards the cosmic microwave background (CMB) \citep[e.g.][]{Komatsu2011Seven-yearInterpretation,PlanckCollaboration2018}.

The ground state of neutral hydrogen atom can produce a signal through a spin-flip transition, which is known as the 21-cm signal. This signal will be a unique signature of EoR \citep[e.g.][]{Madau199721Redshift,Furlanetto2006CosmologyUniverse}. When observed, this 21-cm signal would have redshifted to radio band of the electromagnetic spectrum. Various radio experiments, such as Low Frequency Array\footnote{\url{https://www.astron.nl/telescopes/lofar/}} \citep[LOFAR; e.g.][]{vanHaarlem2013LOFAR:ARray}, Murchison Widefield Array\footnote{\url{https://www.mwatelescope.org/}} \citep[MWA; e.g.][]{tingay13} and the Hydrogen Epoch of reionization Array\footnote{\url{http://reionization.org/}} \citep[HERA; e.g.][]{deboer2017hydrogen}, have been trying to detect this signal. Recently these facilities have provided useful upper limits on the 21-cm power spectrum \citep[e.g.][]{Mertens2020ImprovedLOFAR,Trott2020Deepobservations} that have been used to derive constraints on the properties of reionization \citep[e.g.][]{Ghara2020ConstrainingObservations,Ghara2021Constraining,Mondal2020TightLOFAR,Greig2020ExploringSignal,Greig2020Interpretingobservations}.

The 21-cm signal during the EoR will be highly non-Gaussian and therefore the power spectrum will not give a full statistical characterisation of it \citep[e.g.][]{Mellema2006,2010MNRAS.406.2521I,Watkinson2015TheMoments,Majumdar2018QuantifyingBispectrum,Giri2019Position-dependentReionization}.
In the coming years, the Square Kilometre Array\footnote{\url{https://skatelescope.org}} (SKA) will be built. The low-frequency component of the SKA will be sensitive enough to detect the 21-cm signal produced during EoR and create images of its distribution on the sky \citep{Mellema2013ReionizationArray, Wyithe2015ImagingSKA, Koopmans2015TheArray}. These images contain more information about our Universe as the detection of the signal at different observed frequencies depict the distribution of neutral hydrogen at a given time during the EoR. 

SKA-Low will observe a sequence of such 21-cm images from different redshifts that will constitute a three-dimensional set of data known as a tomographic dataset. The evolution of the 21-cm signal can be seen along the redshift axis. See for example \citet{giri2019tomographic} for more description about tomographic 21-cm images. The reionization process is driven by growing \HII regions, often referred to as bubbles \citep[e.g.][]{Furlanetto2004}. As the sources of ionizing photons reside inside them, observing these bubbles and their evolution will be interesting. Numerous studies have provided various methods to detect and study properties of \HII bubbles \citep[e.g.][]{Datta2007DetectingMaps,Zackrisson2020Bubble6,Gronke2020MeasuringLySpectra}. We can also study the properties of reionization with 21-cm images \citep{Giri2018BubbleTomography,Giri2019NeutralTomography}. However, tomographic images from SKA-Low will be prone to instrumental limitations, such as noise, limited resolution and foreground contamination \citep[e.g.][]{Koopmans2015TheArray,Ghara2016}. In the field of image processing, methods that can classify objects or features in images into meaningful segments are known as `image segmentation' methods. \cite{Giri2018BubbleTomography} implemented an image segmentation method to classify neutral and ionized regions in 21-cm images in the presence of instrumental limitations and demonstrated that key properties of reionization can be derived from such observations.

Artificial intelligence (AI) and deep learning methods are capable of learning patterns in image data and identifying interesting regions. Image segmentation based on AI 
%is a common problem in the field of image processing, aiming to learn visual patterns from an input image and identify objects classes that make up that image. This technology 
is quite popular in the field of data analysis and has been applied to study objects with complex visual form contained in big data \citep{Long2014}. In recent years, several papers made use of machine learning techniques for a range of problems in astrophysics \citep[e.g.][]{Lee2019, Giri2019IdentifyingSurveys, Yoshiura2020, Chen2020} and cosmology \citep[e.g.][]{Jeffrey2020, Sadr2020, Guzman2021}. In the case of reionization, several of these methods are aimed to either remove foreground emission \citep{2019MNRAS.485.2628L, Makinen2020, Villanuevadomingo2020}, emulate reionization simulations \citep[e.g.][]{Kern2017EmulatingHeating,2018MNRAS.475.1213S,Jennings2018,Cohen2020Emulating,Ghara2020ConstrainingObservations} or constrain reionization history \citep[e.g.][]{Shimabukuro2017, Chardin2019, Mangena2020, shimabukuro2020power} and its astrophysical inputs \citep[e.g.][]{Sullivan2018, Gillet2019, Hassan2020}. 

In this work, we present a new approach for the identification of the distribution of \HII regions in 21-cm images using a deep learning method named U-shaped convolutional neural network (U-Net), which is specially designed for image segmentation and feature extraction \citep{Ronneberger2015}. In our case, we adapt this network for processing our image data, which are mock observations of the 21-cm signal during the EoR. The method will segment the images into ionized and neutral regions. We call this framework \texttt{SegU-Net}.

This paper is organised as follows. In \secref{chap:21cmsignal} we present how we generate the simulated data sets used for this work. In \secref{chap:unetfor21cmimages} we describe the design of our neural network, including the error estimation. In \secref{chap:results} we discuss its application to our simulated SKA-Low data sets, considering a range of summary statistics such as the mean ionization fraction, power spectra and topological quantities such as size distributions and Betti numbers. In \secref{chap:other_noise} we test our framework on various instrumental noise levels, and in \secref{chap:c2ray} we test it on a data set produced from a fully numerical reionization simulation. We discuss and summarize our conclusions in \secref{chap:discussion}.

\begin{figure*}
	\includegraphics[scale=0.5]{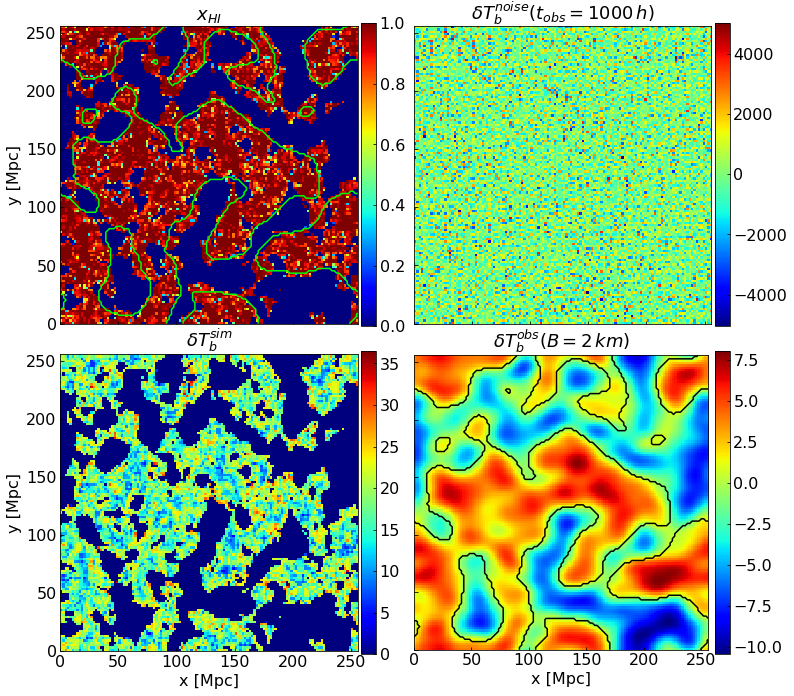}\vskip-2mm
	\caption{\textit{Top left}: the neutral hydrogen fraction at simulation resolution. Green contours indicate the boundary between neutral and ionized regions after reducing the resolution to an observation with a maximum baseline of $B=2\,\rm km$ and matching frequency resolution. \textit{Bottom left}: the 21-cm signal at simulation resolution. \textit{Top right}: The 21-cm signal plus noise realisation at simulation resolution for an observing time of $1000$ hours. To mimic the effect of the lack of a zero baseline, the mean signal has been subtracted. \textit{Bottom right}: The noisy 21-cm image after smoothing to the resolution to an observation with a maximum baseline of $B=2\,\rm km$ and matching frequency resolution. This is an example of a smoothed box slice used during the network training. The solid black line shows the same contour as in the top left panel.}
	\label{fig:train_example}
\end{figure*}

\begin{table}
	\centering
	\caption{The parameters used in this study to model the telescope properties.}
	\label{tab:telescope_param}
	\begin{tabular}{lccccc} % four columns, alignment for each
		\hline
		Parameters & Values \\
		\hline
		%Maximum baseline ($B$) & 2 km  \\
        %\rm observation time ($t_\mathrm{int}$)& 1000 h  \\
        System temperature & $60 (\frac{\nu}{300\mathrm{MHz}})^{-2.55}$ K  \\
		Effective collecting area  & 962 $\mathrm{m}^2$  \\
        %Critical frequency ($\nu_\mathrm{c}$) & 110 MHz \\
        Declination & -30$^\circ$ \\
        observation hour per day & 6 hours \\
        Signal integration time & 10 seconds \\
		\hline
	\end{tabular}
\end{table}
%%%%%%%%%%%%%%%%%%%%%%%%%%%%%%%%%%%%%%%%%%%%%%%%%%%%%%%%%%%%%%%%%%%%%%%%%%%%%%%%%
\section{21-cm signal} \label{chap:21cmsignal}
For any deep learning based method, we need a data set containing a sample of all the possible scenarios, known as the training set. In \secref{sec:simulation}, we describe the reionization simulation code that we use to create the training set. The observable for radio telescopes observing the 21-cm signal is defined in \secref{sec:dTb}. Finally, in \secref{sec:mock_obs} we give the methodology we use to mimic the observations expected with SKA-Low.

\subsection{Reionization simulation}\label{sec:simulation}
To train our network, we require a large set of simulations that represent the 21-cm radio signal for a wide range of redshift during reionization and different assumptions about the astrophysical sources
of ionizing radiation. To do so we employ \texttt{py21cmFAST}, the \texttt{Python} wrapped version of the semi-numerical cosmological simulation code \texttt{21cmFAST} \citep{Mesinger2011,Murray202021cmFAST}. The code computes the evolution of the matter density field using the Zel'dovich approximation \citep{Zeldovich1970}. The ionization field and the corresponding 21-cm differential brightness temperature are then calculated from the matter density distribution based on the excursion set formalism \citep{Furlanetto2004,Mesinger2007EfficientReionization}, which considers a region to be ionized when the fraction of collapsed matter fluctuation exceeds a mass threshold. The ionization fraction $x_{\rm HII} (\pmb{r})$ at a position $\pmb{r}$ is given as, 
\begin{equation}
    x_{\rm HII} (\pmb{r}) = 
    \begin{cases}
        \, 1 & \text{if} \,\,\, f_\mathrm{coll} \geq 1/\zeta\\
        \, 0 & \text{otherwise}
    \end{cases} 
    \label{eq:excursion_set_condition}
\end{equation}
where $\zeta$ is the ionizing efficiency of high redshift galaxies and $f_{\rm coll}(R_{\rm s},\,M_{\rm min})$ is the fraction of collapsed matter within radius $R_{\rm s}$ that can form haloes with mass greater than $M_{\rm min}$. $f_{\rm coll}$ is calculated at every pixel varying $R_{\rm s}$ within 0 and $R_{\rm mfp}$. The maximum value of $f_{\rm coll}$ is used in \autoref{eq:excursion_set_condition}. $R_{\rm mfp}$ implements the effect of a finite mean free path for ionizing photons in the ionized IGM. 

The cosmological parameters considered in this work are based on WMAP 5 years data observation \citep{Komatsu2009} and consistent with \cite{PlanckCollaboration2018} results. We assume a flat $\Lambda$CDM cosmology with the following parameters, $\Omega_{\Lambda}=0.73$, $\Omega_{m}=0.27$, $\Omega_b=0.046$, $ H_0=70\,km\,s^{-1}\mpc^{-1}$, $\sigma_8=0.82$, $ n_s=0.96$.

%%%%%%%%%%%%%%%%%%%%%%%%%%%%%%%%%%%%%%%%%%%%%%%%%%%%%%%%%%%%%%%%%%%%%%%%%%%%%%%%%
\subsection{Differential brightness temperature}\label{sec:dTb}
Radio interferometry based telescopes record the differential brightness temperature $\delta T_\mathrm{b}$ while observing the redshifted 21-cm signal. $\delta T_\mathrm{b}$ depends on position on the sky $\pmb{r}$ and redshift $z$ and can be given as \citep[e.g.][]{Mellema2013ReionizationArray},
\begin{eqnarray}
    \delta T_\mathrm{b} (\pmb{r}, z) \approx 27 x_\mathrm{HI}(\pmb{x}, z) \big(1 + \delta_\mathrm{b} (\pmb{r}, z) \big) \left( \frac{1+z}{10} \right)^\frac{1}{2}
    \left( 1 -\frac{T_\mathrm{CMB}(z)}{T_\mathrm{s}(\pmb{r}, z)} \right)\nonumber\\
    \left(\frac{\Omega_\mathrm{b}}{0.044}\frac{h}{0.7}\right)
    \left(\frac{\Omega_\mathrm{m}}{0.27} \right)^{-\frac{1}{2}} 
    \mathrm{mK} 
    \label{eq:dTb}
\end{eqnarray}
\noindent where $ x_\mathrm{HI}$, $\delta_\mathrm{b}$, $T_\mathrm{CMB}$ and $T_\mathrm{s}$ are neutral fraction, baryon density contrast, CMB temperature and spin temperature respectively. 

Previous studies have shown that our Universe will be heated before reionization begins \citep[e.g.][]{Pritchard200721-cmReionization,Ross2017SimulatingDawn,Ross2019EvaluatingDawn}. Therefore we assume $T_\mathrm{s}\gg T_\mathrm{CMB}$ throughout this work, which is known as the spin saturated approximation and is relevant at lower redshift $ z \lesssim 12$ \citep[e.g.][]{Furlanetto2004, Furlanetto2006Global21cm}. In the spin saturated approximation scenario, the differential brightness signal is always in emission ($\delta T_{\rm b} \geq 0~\mathrm{mK})$ and locations with $\delta T_{\rm b} = 0~\mathrm{mK}$ correspond to \HII regions.

%\SG{\bf SG: Have you included RSD while simulating? 21cmFast has a knob somewhere. I don't remember what is the default. Please find it out. We have to mention about RSD accordingly.} \MB{MB: Not included as by default 21cmFAST does not include RSD see at \url{https://21cmfast.readthedocs.io/en/latest/_modules/py21cmfast/inputs.html?highlight=space percent20distortion#} in class: FlagOptions} 

%%%%%%%%%%%%%%%%%%%%%%%%%%%%%%%%%%%%%%%%%%%%%%%%%%%%%%%%%%%%%%%%%%%%%%%%%%%%%%%%%
\subsection{Mock 21 cm observation} \label{sec:mock_obs}
In order to train \texttt{SegU-Net} for application to actual observations, we need a training set of mock observations. We create these mock observations by simulating the $\delta T_\mathrm{b}$ using the methods described in previous sections and adding instrumental effects, such as the absence of zero baselines, limited resolution and noise. We follow the methods in \cite{Ghara2016} and \cite{Giri2018BubbleTomography} for mimicking the expected effects of SKA1-Low.

We consider a simulation volume of $(\rm 256\, \mpc)^3$ and an intrinsic resolution of $ \Delta x = 2\, {\rm \mpc}$ for simulating the signal. This intrinsic resolution corresponds to an angular aperture of $\Delta\theta=0.777$ arcmin and a frequency depth of $\Delta\nu=0.124\,{\rm MHz}$ along the line of sight at $ z=7$. As an example, in \autoref{fig:train_example}, we show a coeval cube slice of the neutral fraction field and $\delta T_\mathrm{b}$ field in the top left and bottom left panels, respectively. These slices are taken from the epoch when the universe was about 50 per cent ionized. For each $\delta T_\mathrm{b}$ coeval cube, we assume one axis as the line of sight or frequency direction and subtract the mean signal from each frequency channel, such that this could be considered as a sub-volume from the 3D tomographic data set. We consider this simulation as our reference throughout the results analysis in \S\ref{chap:results}, its astrophysical parameters are given in \autoref{tab:refmodel_param}.

We simulate the instrumental noise using the method given in \cite{Giri2018BubbleTomography} and implemented in \texttt{Tools21cm}\footnote{A python package for EoR simulations analysis. \url{https://github.com/sambit-giri/tools21cm}} \citep{Giri2020t2c}. We change the noise seed for each new member of the training set so that the network is trained on different noise realisations and we list our assumed parameters for the telescope setup in \autoref{tab:telescope_param}. In the top right panel of \autoref{fig:train_example}, we show a slice from the simulated noise cube produced from 1000 hours of observation with SKA1-Low at simulation resolution. When we add this noise to our simulated signal at the simulation resolution, we cannot discern any feature of the signal as the noise is several orders of magnitude higher than the signal. Therefore we reduce the resolution of the noisy signal in the field-of-view direction by smoothing with a Gaussian kernel with full-width at half maximum (FWHM) of $\lambda_0 (1+z)/B$, where $B$ is the maximum baseline. For example, $B=2\,{\rm  km}$ corresponds to a resolution of 2.905 arcmins at redshift $z\approx 7$ and 3.631 arcmins at redshift $z\approx 9$ respectively. In the frequency direction we reduce the resolution by convolving with a top-hat bandwidth filter of a width matching the FWHM of the angular smoothing in comoving units. This width corresponds to 0.462 MHz at redshift $z\approx 7$ and 0.551 MHz at redshift $z\approx 9$ respectively. In the bottom right panel of \autoref{fig:train_example} we show a slice from our noisy signal at this reduced resolution. At this resolution, the smallest \HII regions seen in the top left panel of ~\autoref{fig:train_example} can no longer be discerned. However, we can still identify the larger \HII regions. 

To illustrate what we can achieve with these images, we apply the same smoothing to the neutral fraction field and apply a threshold of $x_{\rm th} = 0.5$ to label neutral/ionized regions. We refer to the smoothed and then binarised neutral fraction field as the \textit{ground truth}. We use this field to compare the accuracy of the recovered binary field throughout our paper. We want to point out that this is different from the ground-truth of the original reionization simulation as the limited resolution of the radio telescope will limit the observation of small scale features. Then, we over-plot the boundaries of these ionized regions, the neutral fraction slice and signal slice in top-left and bottom-right panels of \autoref{fig:train_example} respectively.

\begin{table}
	\centering
	\caption{Astrophysical parameters used for our fiducial simulation.}\vskip-2mm
	\label{tab:refmodel_param}
	\begin{tabular}{lccccc} % four columns, alignment for each
		\hline
		Parameters & Values \\
		\hline
        $ \zeta$       & $39.204$ \\
		$ R_{\rm mfp}$ & $12.861$ \mpc  \\
        $ T^{\rm min}_{\rm vir}$ & $3.46\times 10^4$ K \\%$4.594$ K\\
		\hline
	\end{tabular}
\end{table}

\subsubsection{Training and testing set}
For our training set, we randomly sample the astrophysical simulation parameters by a normal distribution, such that the ionizing efficiency of high-redshift galaxies $\zeta$ is sampled with $\mathcal{N}\sim(52.5, 20)$, the mean free path of ionizing photons $ R_{\rm mfp}$ with $\mathcal{N}\sim(12.5\,\mathrm{\mpc}, 5\,\mathrm{\mpc})$ and the (logarithmically-spaced) minimum virial temperature for halos to host star-forming galaxies $T^{\rm min}_{\rm vir}$ with $\mathcal{N}\sim(4.65, 0.5)$. The choice of these values is such that for a majority of the samples most of the reionization history ($x^{\rm V}_{\rm HI}$ from 0.9 to 0.1) falls within the redshift interval 9 to 7. The redshift is randomly sampled with a uniform distribution $\mathcal{U} \sim [7,\,9]$. The initial conditions of the cosmological density field are changed for each simulation. This helps us avoid the impact of cosmic variance on our trained model. With the list of all the parameter values, we produce 10,000 mock observations of the 21-cm signal. Out of these mock observations, we use 15 per cent as the so-called network validation set. This validation set is used during the training method to provides an unbiased evaluation of the network model fit. 

Eventually, we will use \texttt{SegU-Net} on actual 21-cm image observations. Here we rely on an additional 300 mock observations as the testing or prediction set. Just as for the training set, the parameter values are randomly chosen. We call this the `random' testing set. The training process is blind to the prediction set. Apart from the above testing set, we create an additional simulation with fixed values of astrophysical parameters (given in \autoref{tab:refmodel_param}). We have chosen these values such that between $z = \rm 9$ and 7 reionization proceeds from $x^{\rm V}_{\rm HI}\approx 0.9$ to 0.1. We call this set the `fiducial' testing set. Since the signal evolves as reionization progresses in this testing set, it better mimics the upcoming 21-cm observations. With this testing set, we will test \texttt{SegU-Net}'s capability to capture the evolution of structures and recover the binary field from untrained data in \secref{chap:results}.
\begin{figure}
	\centering
	\includegraphics[width=\columnwidth]{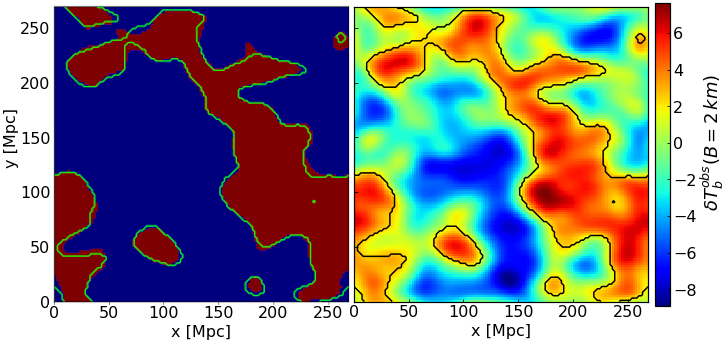}\vskip-2mm
	\caption{\textit{Right panel}: An example of a smoothed cube slice from the \ctworay simulation on the right employed to test the stability and reliability of the network. This slice is for $ z=8.06$ and  corresponds to a volume-averaged neutral fraction of $ x^{\rm V}_{\rm HI}=0.38$. As for the training set, at simulation resolution, we subtracted the mean signal in the frequency direction from the differential brightness temperature. We then added simulated instrumental noise for the observed time of 1000 hours and smoothed the signal with the same baseline as SKA1-Low. \textit{Left panel}: the binary field recovered with our neural network. In red/blue, the prediction performed with our network and the green contour shows the boundary between neutral and ionized region. The same contour is shown with a solid black line on the right panel for comparison.}
	\label{fig:c2ray_slice}
\end{figure}
\subsubsection{Fully-numerical simulations testing set} \label{sec:c2ray_method}
To train \texttt{SegU-Net}, we relied on \texttt{21cmFAST} for creating the training set. However, our Universe may not exactly be described by this semi-numerical code. If our neural network has learnt to find structures in \texttt{21cmFAST} simulations only, then we cannot use it for SKA observations. To ensure that the neural network is not over-fitted, we consider a different reionization simulation code to build the mock observations.

We first simulate the matter density field and track the evolution of cosmic structures by using the \texttt{CUBEP$^3$M} $N$-body code \citep{Harnois-Deraps2013}. The simulation is carried out in a volume of $(349 {\rm \mpc})^3$ with 64 billion particles. Dark matter haloes down to a mass of $10^9 M_\odot$ is found at various redshift using the spherical average halo-finder \citep{Watson2013TheAges}, meanwhile haloes with masses between $10^8$ and $10^9\,M_\odot$ are implemented with a sub-grid method \citep{Ahn2014}. We use the same cosmology that is given in \secref{sec:simulation}.

\begin{figure*}
	\includegraphics[width=\textwidth]{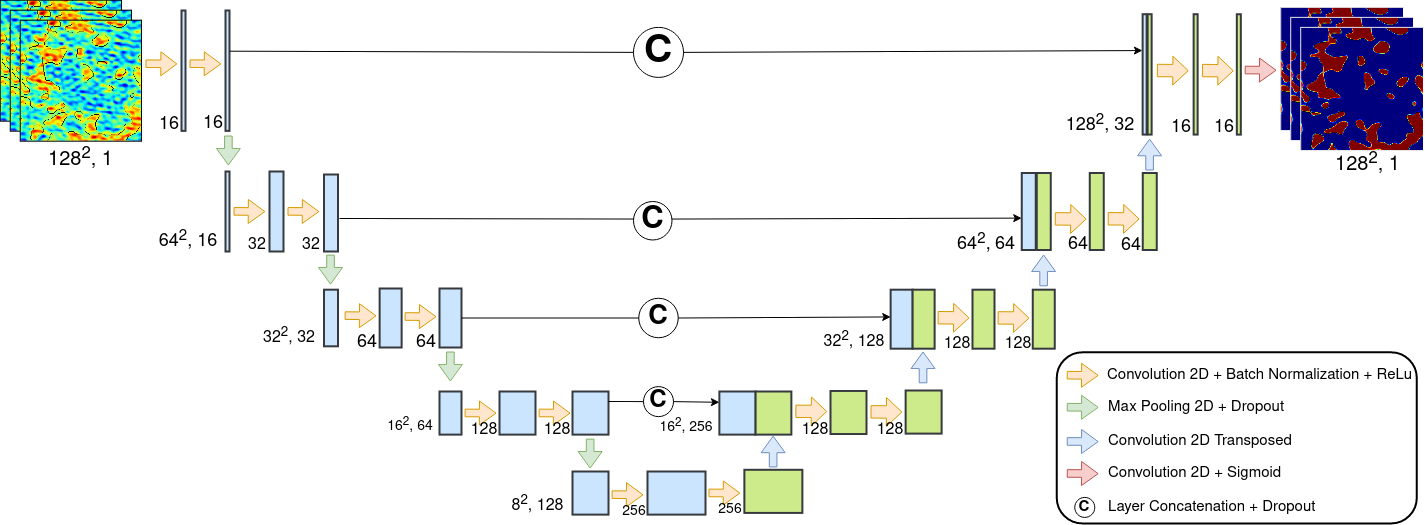}\vskip-2mm
	\caption{Schematic representation of \texttt{SegU-Net} network architecture. The orange arrow indicates a 2D convolutional layer, followed by batch normalization and ReLU activation. Pooling operations followed by dropout layer are indicated with green arrows. The blue arrow indicates an up-sampling layer by transposed 2D convolutional layer and with a red arrow the closing layer, a 2D convolution followed by a sigmoid activation function. The descending path on the left side divides the resolution of the image after each pooling operation and doubles the channel dimension after each convolution. On the other hand, the expansion path doubles the spatial dimension at each up-sampling operation and decreases the channel dimension after concatenation with its counterpart layer in the descending path.}
	\label{fig:architecture}
\end{figure*}

We then employ the \ctworay radiative transfer (RT) code \citep{Mellema2006} to simulate the cosmic reionization. \ctworay requires the matter density field in a 3D grid. Therefore the distribution $N$-body particles are put in 3D grids with a smoothed particle hydrodynamic method \citep[e.g.][]{Shapiro1996,Mao2019}.
This grid has spatial resolution of $ \Delta x = 2.1$ \mpc~and a $166^3$ mesh-grid. Sources ionizing photon production rate per unit time is proportional to the mass of the hosting halo $M_{\rm halo}$ such that.
\begin{equation}
    \dot{N}_{\gamma} = f_{\gamma} \frac{M_{\rm halo}\,\Omega_b}{\Delta t_s\,m_p\,\Omega_m}
\end{equation}
where $m_p$ is the proton mas and $\Delta t_s=11.53\,\rm Myr$ is the stars lifetime. The efficiency factor of sources is defined as $f_{\gamma} = f_{\star}\, f_{\rm esc}\, N_i$ where $f_{\star}$ is the star formation efficiency, $f\rm_{esc}$ is the photons escape fraction and $N_i$ is the stars ionizing photon production efficiency per stellar atom. The efficiency factor for halos with masses $M_{\rm halo}<10^9 M_{\odot}$ is set to $f_{\gamma}=2$. For the the lower mass halos it is initially set to $f_{\gamma}=8.2$. When their environment becomes ionized (above 10 percent), their efficiency is reduced to $f_{\gamma}=2$ to account for radiative feedback. \ctworay outputs the hydrogen ionization field at a time interval of 11.5 million years. For more details on the RT and $N$-body simulations methods, see \citet{Iliev2012CanSources} and \citet{Bianco2021}.

We derive the differential brightness temperature $\delta T_{\rm b}$ from the ionization field and the density using \autoref{eq:dTb}. We select four outputs, which are at redshifts $z=7.96,\,8.06,\,8.17,\,8.28,\,8.40,\,8.52,\,8.64$, corresponding to a volume averaged neutral fraction of $x^{\rm V}_{\rm HI} = 0.17,\,0.29,\,0.42,\,0.57,\,0.70,\,0.81,\,0.90$, respectively. The simulated $\delta T_{\rm b}$ from these epochs are converted into mock observations using the procedure outlines in \secref{sec:mock_obs}. We use these mock observations as a testing set.

The right panel of \autoref{fig:c2ray_slice} shows a slice of the calculated $\delta T_{\rm b}$ for redshift $z=8.06$ ($x^{\rm V}_{\rm HI}=0.38$ at simulation resolution). Similar to the bottom right panel of \autoref{fig:train_example}, we add the instrumental noise corresponding to a 1000~h observation and smooth the signal to a resolution corresponding to a maximum baseline $ B=2\,{\rm km}$. The black contours correspond to the boundary between neutral and ionized regions. These boundaries are derived from the simulated neutral fraction field at the same resolution as the $\delta T_{\rm b}$ data set.

%%%%%%%%%%%%%%%%%%%%%%%%%%%%%%%%%%%%%%%%%%%%%%%%%%%%%%%%%%%%%%%%%%%%%%%%%%%%%%%%%
\section{U-Net for 21-cm image segmentation}\label{chap:unetfor21cmimages}
Here we describe our machine learning method for identifying ionized and neutral regions in noisy 21-cm images and our approach to estimate the uncertainty of its results in \secref{sec:ournet} and \secref{sec:confidence_interval} respectively.

\begin{figure*}
	\includegraphics[width=\textwidth]{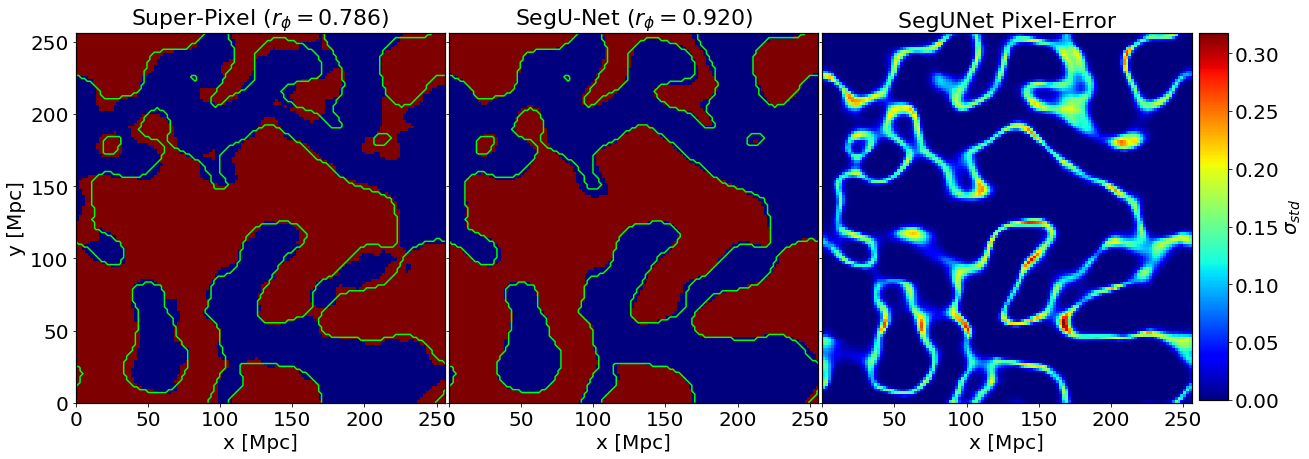}\vskip-2mm
	\caption{Slice comparison of the binary field, in blue ionized regions and in red neutral. \textit{Left panel}: binary field recovered by the Super-Pixel method. \textit{Central panel}: binary field recovered by our neural network. Green lines indicate the true separation between ionized/neutral regions, derived from a smoothed version of the simulated neutral hydrogen distribution. \textit{Right panel}: the per-pixel error as calculated by \texttt{SegU-Net}. The color-bar indicates the intensity of the network uncertainty.}
	\label{fig:visual_comparison}
\end{figure*}

%%%%%%%%%%%%%%%%%%%%%%%%%%%%%%%%%%%%%%%%%%%%%%%%%%%%%%%%%%%%%%%%%%%%%%%%%%%%%%%%%
\subsection{Our network, \texttt{SegU-Net}}\label{sec:ournet}
Our segmentation network\footnote{\url{https://github.com/micbia/SegU-Net}} is based on the U-Net framework first introduced by \cite{Ronneberger2015}. U-Net consists of two likewise symmetric paths, an encoder operator that contracts the image and a decoder operator that expands the extracted features. The encoder corresponds to a classical convolutional neural network (CNN). This CNN aims to reduce the size of the input image in such a way that only information of the most interesting features remains. A series of concatenated convolution operations (layers) returns a low dimensional latent space (or latent vector) that contains information about these extracted features. In Appendix \ref{app:hidden_layer} we provide a visual representation of the low dimensional latent space for the example case of a sphere. We show a schematic representation of the U-Net in \autoref{fig:architecture}. The left part of the U-shape in the diagram and the bottom layer represents the encoder and the low dimensional latent space respectively. For a detailed discussion of CNNs, we refer the reader to \cite{Mehta2019}, and for examples of employing CNNs to infer cosmological and astrophysical parameters in the context of reionization to \cite{Gillet2019} and \cite{Hassan2020}. 
In our case, the information in the latent space (or latent vector) of U-Net (bottom layer) is expanded by a decoder into a binary map of the same size as the input image. The right part of the U-shape of the diagram in \autoref{fig:architecture} represents the decoder. The decoder gradually increases the spatial resolution of the latent vector with an up-sampling operation (transposed convolution) until we obtain the same dimension of the input image. After each up-sampling step, the output is combined with the corresponding encoder layer with the same dimension. We illustrate this further in Appendix \ref{app:concatenation_example} with an example.

Even though each of our image data sets is 3D, \texttt{SegU-Net} is trained on 2D slices. We identify structures in 3D image data by running on every slice along the third axis. Tests show that the method is not sensitive to the choice of the third axis. When compared to a neural network trained on 3D data, we found that our approach is computationally less expensive without loss of accuracy. Therefore the U-Net architecture described in this work is only applied to 2D image data.

The structure of the encoder layers consists of two convolutional blocks followed by a 2D max-pooling layer (\texttt{MaxPool}) of size 2x2 and a 5 per cent rate dropout layer (\texttt{Drop}). This regularization technique randomly shuts down a portion of the layer neurons to avoid over-fitting \citep{Hinton2012, Srivastava2014}. The convolutional block (\texttt{ConvBlock}) consists of a 2D convolution layer (\texttt{Conv2D}) with 3x3 kernel size. We add a layer that normalizes the previous input layer over the batch sample to avoid over-fitting (\texttt{BN}) \citep{Ioffe2015} and as an activation function we employ a Rectified Linear Unit (\texttt{ReLU}) activator \citep{Jarrett2009, Glorot2011}, \texttt{ConvBlock=Conv2D+BN+ReLU}. This layer structure is repeated for a total of four levels (\texttt{Encoder-Level}). At each step, the dimension of the input image is halved by the max-pooling operation. The number of feature channels is doubled by the convolutional layer, \texttt{Encoder-Level=2*ConvBlock+MaxPool+Drop}. The decoder structure is somewhat similar to the encoder. We replace the pooling operation with a transposed 2D convolution (\texttt{TConv2D}) \citep{Dumoulin2016, Zeiler2013}, that has an opposite scaling effect on the resolution and channel size. This layer output is then concatenated (\texttt{CC}) with the corresponding encoder level to preserve the features extracted in the contracting path. This step is followed by a dropout layer and two convolutional blocks, \texttt{Decoder-Level=TConv2D+CC+Drop+2*ConvBlock}. The final output consists of a 2D convolutional layer followed by a sigmoid activation. Our network has a total of 23 2D convolutional layers distributed on four down- and up-sampling scaling levels and a bottom layer, for a total of approximately 2.5 million trainable parameters. In \autoref{fig:architecture}, we show our best performing network and label the shape of the output from each intermediate hidden layer of this network. More details are provided in Appendix \ref{app:hidden_layer} and \ref{app:concatenation_example}. 

During our training process, the hyperparameters of the network are learnt by minimizing a loss function. We employ the balanced cross-entropy (BCE) \citep{Salehi2017}, 
\begin{equation}\label{eq:bce}
    \rm \mathcal{L}(y, \hat{y}) = -\frac{1}{N} \sum^{N}_{i=0} \left(\beta \, y_i \,log_{10}(\hat{y}_i) + (1-\beta)(1-y_i)log_{10}(1-\hat{y}_i) \right)
\end{equation}
where $ y_i \in \{0;1\}$ is the pixel-wise ground truth, $\hat{y_i}$ the predicted value, $ N$ the batch size, which is our case is of size 32 and the parameter $\beta = \sum^{N}_{i=0} y_i$ is the average volume neutral fraction of the batch. In our context, at early/late stage of reionization the statistical weight of the ionized/neutral pixels are under-represented. This situation is known in data science as a problem affected by \textit{``class unbalanced'' data}. To deal with this we use the above loss function which has been shown to be well suited for segmentation problems that are affected by class unbalanced data \citep{Cui2019}. We further used the Adaptive Moment Estimator Adam \citep{Kingma2014}, an optimized stochastic gradient descent algorithm for error minimization. The initial learning rate, the step size of the rate of convergence that minimizes the loss function, is set to $10^{-3}$. We trained the network using 2 GPUs, and it took approximately 1,500 wall clock hours.

%%%%%%%%%%%%%%%%%%%%%%%%%%%%%%%%%%%%%%%%%%%%%%%%%%%%%%%%%%%%%%%%%%%%%%%%%%%%%%%%%
\subsection{Uncertainty estimation on \texttt{SegU-Net}}\label{sec:confidence_interval}
One of the main drawbacks of machine learning is that it is unable to quantify uncertainties and confidence intervals for its predictions, and only recently attempts have been made \citep{Charnock2020Bayesian, Hortua2020} to include error estimation. However, this has not yet been generally implemented for U-Nets. Additionally, if not well optimized, neural networks are prone to over-fitting and tend to be biased. Therefore, we have developed an error estimation procedure to be used during the prediction process. This procedure gives our network additional power by providing a pixel by pixel error map.

Image manipulations, such as zooming, shifting along an axis, flipping axes and rotation along an axis, are commonly performed on 2D or 3D image training data to increase the number of samples \citep{Simonyan2015, Szegedy2015}. This technique is known as time-test augmentation (TTA) of data \citep{Perez2017, Wang2020}. Here we use this approach to estimate the error on the final result. We perform several copies ($\sim 100$) of the same sample during the prediction process through image manipulations. These manipulated copies are then independently processed by \texttt{SegU-Net}. Each of the recovered binary fields is transformed back. We calculate the final result as the average of these fields and the per-pixel standard deviation to estimate the error for each pixel.

An example of the pixel per pixel error map can be seen in \autoref{fig:visual_comparison} (right-most panel). We will discuss this figure further in \secref{sec:visualcomparison}. This simple method provides our neural network with an uncertainty estimation for each labelled pixel.

%%%%%%%%%%%%%%%%%%%%%%%%%%%%%%%%%%%%%%%%%%%%%%%%%%%%%%%%%%%%%%%%%%%%%%%%%%%%%%%%%
\begin{figure*}
	\includegraphics[width=\textwidth]{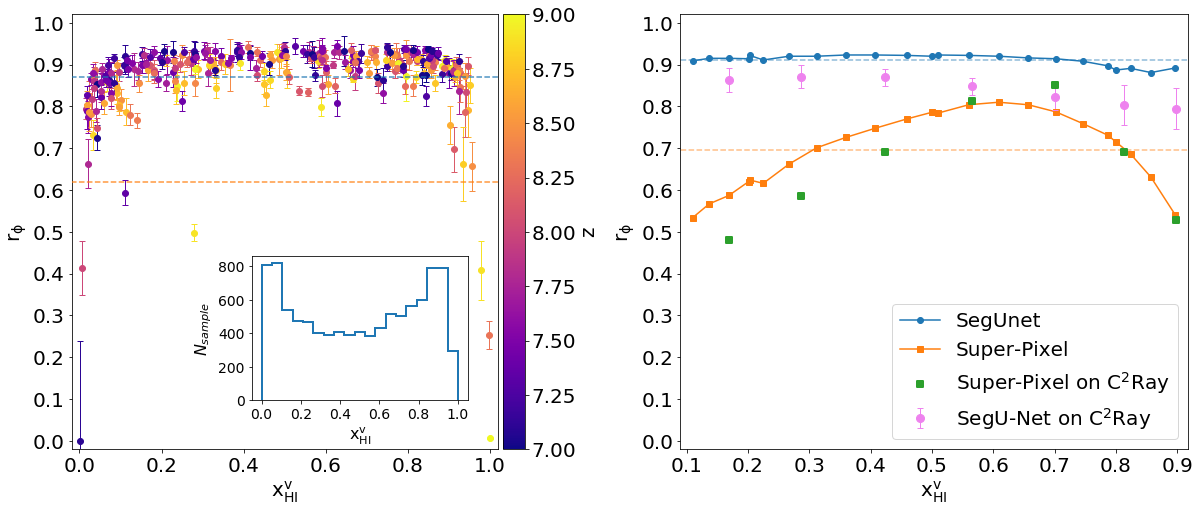}\vskip-2mm
	\caption{\textit{Left panel}: the Matthews correlation coefficient $ r_{\phi}$ of the recovered binary field for the prediction set, against its volume-averaged neutral fraction. Error-bar indicates the network confidence interval, and colours indicate the redshift of the simulated coeval cube. On the inset panel, we show the distribution of the training set (blue histogram) against the volume average neutral fraction. \textit{Right panel}: comparison of the same correlation coefficient for recovery performed on the fiducial simulation with our neural network (blue circle line) and the Super-Pixel method (orange square line). We also include the result from the test on the \ctworay simulation, from left to right, redshift $z=7.96,\,8.06,\,8.17,\,8.28,\,8.40,\,8.52,\,8.64$ corresponding to a volume-averaged neutral fractions of $ x^{\rm V}_{\rm HI} = 0.17,\,0.29,\,0.42,\,0.57,\,0.70,\,0.81,\,0.90$. The violet dots with relative confidence intervals are predictions performed with SegU-Net for these cases and the green squares are the corresponding results from the Super-Pixel method. Horizontal dashed lines in both panels indicate the overall average $r_\phi$ coefficient for the entire data set and the fiducial simulation respectively, in blue for \texttt{SegU-Net} and orange for Super-Pixel method.} \label{fig:phi_coef}
\end{figure*}

\section{Results}\label{chap:results}
Once the network is trained, we want to estimate how well it recovers the binary field from noisy 21-cm images. To do so, we include in our analysis the state-of-the-art Super-Pixel method presented in \cite{Giri2018BubbleTomography}. The Super-Pixel method is based on an advanced image processing technique called the Simple Linear Iterative Clustering (SLIC) \citep{Achanta2012SLICMethods}. SLIC groups similar pixels in images into \textit{``super-pixels''}. These Super-Pixels are then classified into neutral and ionized ones to get the final map containing the identified features. In previous studies, this method has been shown to be superior compared to other methods, such as putting a simple threshold to the mean signal \citep[used in e.g.][]{Kakiichi2017}, the k-means method \citep[used in e.g.][]{Giri2018BubbleTomography} or the maximum deviation method \citep[used in e.g.][]{Gazagnes2021Inferring}. The Super-Pixel method proves to be quite efficient in recovering the binary fields from noisy 21-cm images. The summary statistics extracted from those are accurately reproducing the ones obtained using the simulation data sets. As shown by \cite{Giri2018BubbleTomography}, the choice for the number of super-pixels depends on the simulation box size and resolution. In our case, we tested for a few values between 500 and 7000. We noticed that above the value of 5000, the algorithm becomes more computationally expensive without yielding a substantial increase in the segmentation accuracy. Hence we employ 5000 super-pixels. 
%\SG{[SG: Sorry, I did not notice before submission. k-means method is introduced in \citep{Giri2018BubbleTomography}. \citet{Kakiichi2017} puts a threshold at the mean of the signal.]} 

%\SG{SG: The number of superpixels depend on the boxsize and resolution. In the superpixels paper we gave an ansatz to get this number. One can also do a test by increasing the number. At a certain value there will not be any more improvement in the result. The latter method is more robust. I do not remember which one we used to get 5000 for this work. From the current text it seems that 5000 is a magic number which is not true.}\MB{MB: I changed the text. I explained better why we chose 5000 super-pixels.}

\begin{figure*}
	\includegraphics[width=\textwidth]{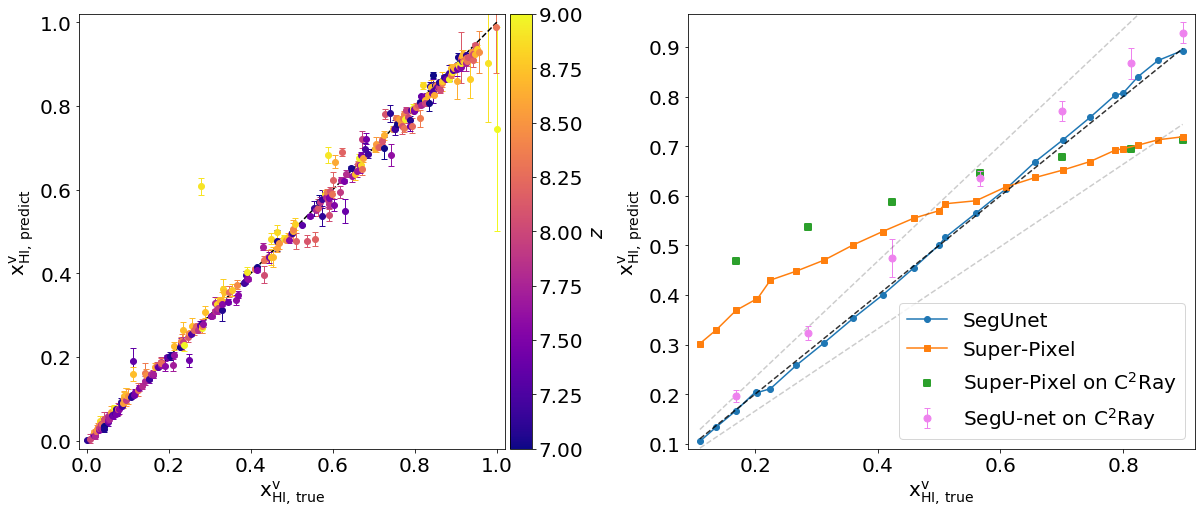}\vskip-2mm
	\caption{\textit{Left panel}: Comparison of the simulated neutral fraction against the recovered one. Error-bar and color-bar are the same as \autoref{fig:phi_coef}. \textit{Right panel}: The same comparison for the `fiducial' simulation. We also include the results from \ctworay simulation. The redshift of \ctworay simulation are $z=7.96,\,8.06,\,8.17,\,8.28,\,8.40,\,8.52,\,8.64$ corresponding to a volume averaged neutral fraction of $ x^{\rm v}_{\rm HI} = 0.17,\,0.29,\,0.42,\,0.57,\,0.70,\,0.81,\,0.90$. The violet dots with relative confidence interval are predictions performed with \texttt{SegU-Net} and green squares are with the Super-Pixel method. } \label{fig:corr}
\end{figure*}
\subsection{Visual comparison}\label{sec:visualcomparison}
To start, we show a visual comparison of slices in \autoref{fig:visual_comparison}. We compare the predicted binary field recovered by the Super-Pixel method (left-most panel) and \texttt{SegU-Net} (central panel) with the ground truth (green contours in both panels). As explained in \secref{sec:mock_obs} the ground truth is the boundary of ionized regions extracted from the simulation neutral fraction field at the same resolution by putting a threshold of 0.5. The red and blue pixels represent neutral and ionized pixels, respectively. In the right-most panel, we show the pixel-error estimated from \texttt{SegU-Net} with a colour bar. The error is determined by calculating the standard deviation of the same pixel from the different version of the same mock observation produced with TTA (see \secref{sec:confidence_interval}).

\texttt{SegU-Net} shows better precision in recovering shapes of the ionized regions compared to the Super-Pixel method. As expected, most of the network uncertainty is located at the boundaries of neutral regions or between two large ionized bubbles when these are percolating, and the gap is getting narrower. This uncertainty has a direct bearing on small neutral islands of a few Mpc scale, residing in vast ionized regions. Moreover, larger uncertainties, $\sigma_{\rm std}\geq 0.25$ are located around narrow ionized regions protruding into large neutral regions (e.g. in \autoref{fig:visual_comparison} right-most panel, at coordinates $x\rm\sim140\,\mpc$ and $y\rm\sim125\,\mpc$). This behaviour suggests that the uncertainty mainly depends on the contrast between the local neutral and ionized regions. The network selects regions in the image based on the largest gradient in the 21-cm signal intensities to recover the binary field. Therefore, we expect larger uncertainties for reionization scenarios in which the contrast in the 21-cm intensities are relatively small.

\begin{table}
    \caption{Summary of the Matthews correlation coefficient score (in per cent) of our two test sets for the two feature identification methods.}\label{tab:mcc_table}
    \begin{tabular}{l|ll|ll|}
        & \multicolumn{2}{l|}{\texttt{SegU-Net}} & \multicolumn{2}{l|}{Super-Pixel}\\\hline
        \multicolumn{1}{|l|}{redshift}  & random set    & fiducial  & random set & fiducial\\\hline
        \multicolumn{1}{|l|}{$z\rm\leq7.75$}   & 88.9\%  & 91.7\%    & 63.7\%  & 62.6\% \\ \hline
        \multicolumn{1}{|l|}{$z\rm\geq8.25$} & 85.3\%              & 90.1\%                 & 60.7\%          & 71.8\%              \\ \hline
        \multicolumn{1}{|l|}{${\rm 7}\leq z \rm\leq 9$}     & 87.1\%              & 91.2\%                 & 62.0\%          & 69.5\%              \\ \hline
        \end{tabular}
\end{table}

\subsection{Correlation coefficient}\label{sec:mcc}
To compare the predicted ionized fields from the 21-cm images mathematically, we use the Matthews correlation coefficient (MCC) (also known as $ r_{\phi}$ coefficient) defined as:
\begin{equation}
    r_{\phi} = \frac{N_{\rm TP}\cdot N_{\rm TN} - N_{\rm FP}\cdot N_{\rm FN}}{\sqrt{(N_{\rm TP}+N_{\rm FP})(N_{\rm TP}+N_{\rm FN})(N_{\rm TN}+N_{\rm FP})(N_{\rm TN}+N_{\rm FN})}} \ ,
\end{equation}
where $N_{\rm TP}$ and $N_{\rm TN}$ are the total numbers of neutral and ionized pixels recovered correctly, respectively. $N_{\rm FP}$ is the total numbers of pixels incorrectly guessed as neutral and $N_{\rm FN}$ is the total numbers of pixels incorrectly guessed as ionized. In our case, a positive/negative result corresponds to the neutral/ionized case since the quantity 1 in our binary fields indicates the neutral condition and 0 the ionized. Thus, MCC is a useful metric to correlate binary fields.

In \autoref{fig:phi_coef} we show the MCC estimated from the fields segmented into ionized and neutral regions in our testing sets. In the left panel, we provide a scatter plot of MCC values against the reionization history ($x_\mathrm{HI}^\mathrm{v}$) for the `random' testing set. We indicate the redshift of the realization by the colour of the points and respective confidence interval with an error bar. We show the number of samples in our training set at a different neutral fraction in an inset panel. After a first attempt, we realized that to overcome the unbalanced class problem requires a better representation of the early ($ x^{\rm v}_{\rm HI} \approx 1$) and late stages of reionization ($ x^{\rm v}_{\rm HI} \approx 0$). For this reason, we increased the number of training samples for these stages. Therefore the distribution of samples against neutral fraction has a bimodal shape with peaks at approximately $ x^{\rm v}_{\rm HI} \approx 0.1$ and $ 0.9$. 

As a result, the $ r_{\phi}$ value for the overall prediction data set (\autoref{fig:phi_coef}, left panel) is about 87 per cent for \texttt{SegU-Net} (blue dashed line) and 62 per cent in the case of the Super-Pixel method (orange dashed line). The noise level increases with redshift. Therefore the score is slightly less accurate for redshift $ z\geq8.25$ with an 85 per cent accuracy, meanwhile higher for lower redshift $ z \leq 7.75$ with 88 per cent. In the future, we consider increasing the proportion of the training data with high redshift to decrease this performance dissimilarity. The same trend is present in the case of the Super-Pixel method, with an accuracy of 60 per cent and 63 per cent, respectively.

In the right panel of \autoref{fig:phi_coef}, we compare the MCC values from \texttt{SegU-Net} (blue line with circles) with that from the Super-Pixel method (orange line with squares) for our `fiducial' simulation. As we already know from \cite{Giri2018BubbleTomography}, the Super-Pixel method performs best for $ x^{\rm V}_{\rm HI} \approx 0.5$ and deteriorates towards earlier and later stages of reionization. The reason for this behaviour is that during these stages, structures are usually smaller and, therefore, more difficult to identify. With \texttt{SegU-Net} we are able to overcome this problem by employing a specifically designed BCE loss function (\autoref{eq:bce}) during the validation process after each training epoch. Therefore, the average $ r_{\phi}$ value for the `fiducial' simulation is about 91 per cent for \texttt{SegU-Net} (blue dashed line) and 70 per cent in the case of the Super-Pixel method (orange dashed line). In \autoref{tab:mcc_table}, we summarise the $r_{\phi}$ score for the two test sets.

\begin{figure*}
	\centering
	\includegraphics[width=\textwidth]{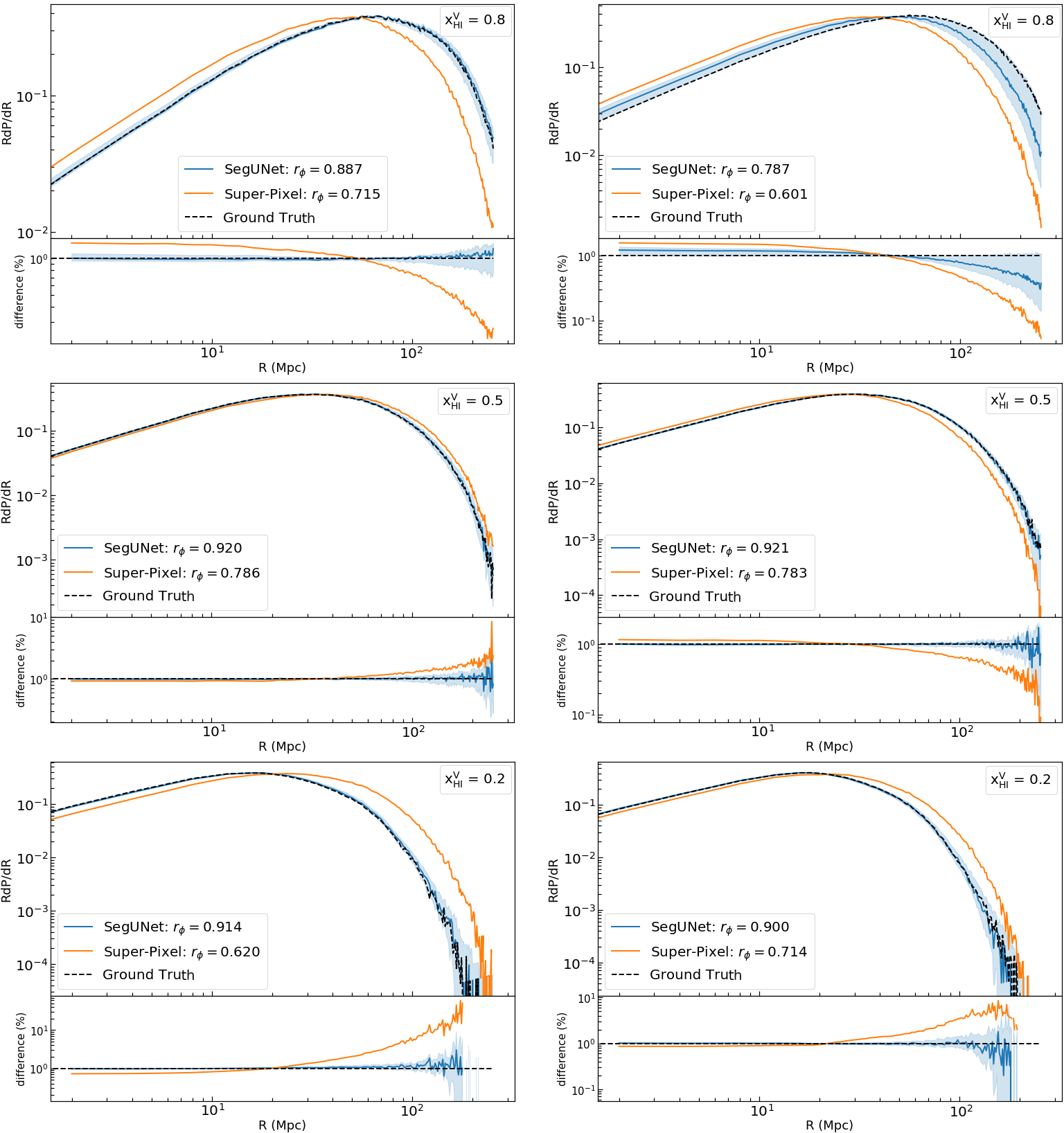}
	\caption{\textit{Left column}: the size distribution of neutral regions (ISD). \textit{Right column}: the size distribution of ionized region (BSD). Rows from top to bottom represents early ($ x^{\rm v}_{\rm HI} = 0.8$), middle ($ x^{\rm v}_{\rm HI} = 0.5$) and late ($ x^{\rm v}_{\rm HI} = 0.2$) stages of reionization respectively.
	On each panel, we show the size distributions from the binary fields of the `fiducial' simulation recovered by \texttt{SegU-Net} (blue line) and its respective confidence interval (blue shadow). Black dashed lines and orange lines give the size distributions of the ground truth and binary field recovered by the Super-Pixel method. At the bottom of each size distribution panel, we show the relative deviation from the true binary field distribution.}
	\label{fig:bs_comparison}
\end{figure*}

\subsection{Average neutral fraction} \label{sec:corr}
After identifying the ionized regions, we can determine the volume-averaged neutral fraction $x^{\rm v}_{\rm HI}$, which quantifies the reionization history. In \autoref{fig:corr} we show the volume-averaged neutral fraction $x^{\rm v}_{\rm HI,\,predicted}$ as calculated from the recovered binary fields extracted by the two methods. In the left panel we show the $x^{\rm v}_{\rm HI,\,predicted}$ from the \texttt{SegU-Net} outputs against the true volume-averaged neutral fraction $x^{\rm v}_{\rm HI,\,true}$ for our `random' testing set. The colour of the points indicates the redshifts. The black dashed line indicates $x^{\rm v}_{\rm HI,\,predicted}=x^{\rm v}_{\rm HI,\,true}$. Except for a few points, all the points lie on or near the black dashed line.

In the right panel of \autoref{fig:corr} we compare the results of $x^{\rm V}_{\rm HI,\,predicted}$ derived with the Super-Pixel method (orange line with squares) and \texttt{SegU-Net} (blue line with circles) for our `fiducial' simulation. Again, the black dashed line represent $x^{\rm V}_{\rm HI,\,predicted}=x^{\rm V}_{\rm HI,\,true}$. In the case of our neural network all results lie within the half standard deviation ($0.5$-$\sigma$) of the true value (gray dashed lines). With Super-Pixel method, this is true only from $x^{\rm V}_{\rm HI} \approx 0.5$ to 0.85. The recovered neutral fraction is either underestimated at $x^{\rm V}_{\rm HI} > 0.6$ or largely overestimate for $x^{\rm V}_{\rm HI} < 0.4$.

\begin{figure}
	\centering
	\includegraphics[width=\columnwidth]{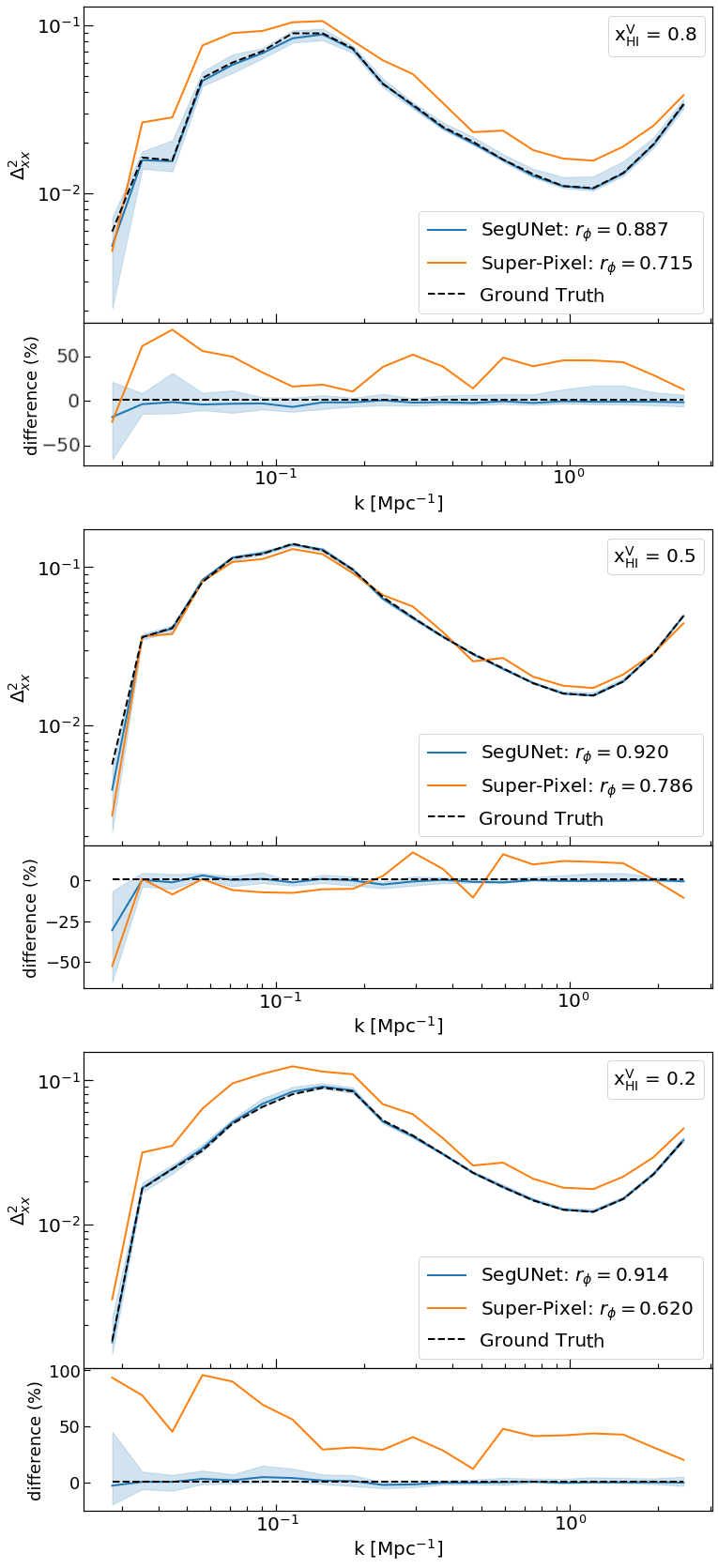}
	\caption{Dimensionless power spectra of the neutral field from the fiducial simulation as recovered by our network (blue line) and its respective confidence interval (blue shadow). Compared at early, middle and late stage of reionization (from top to bottom $ x^{\rm V}_{\rm HI} = 0.8,\,0.5,\,0.2$) with the same quantity derived from the ground truth (black dashed line) and the Super-Pixel method (orange line). At the bottom of each panel, we show the relative difference compared to the ground truth for both cases, the network and Super-Pixel method.}
	\label{fig:pk_comparison}
\end{figure}

\begin{figure}
	\centering
	\includegraphics[width=\columnwidth]{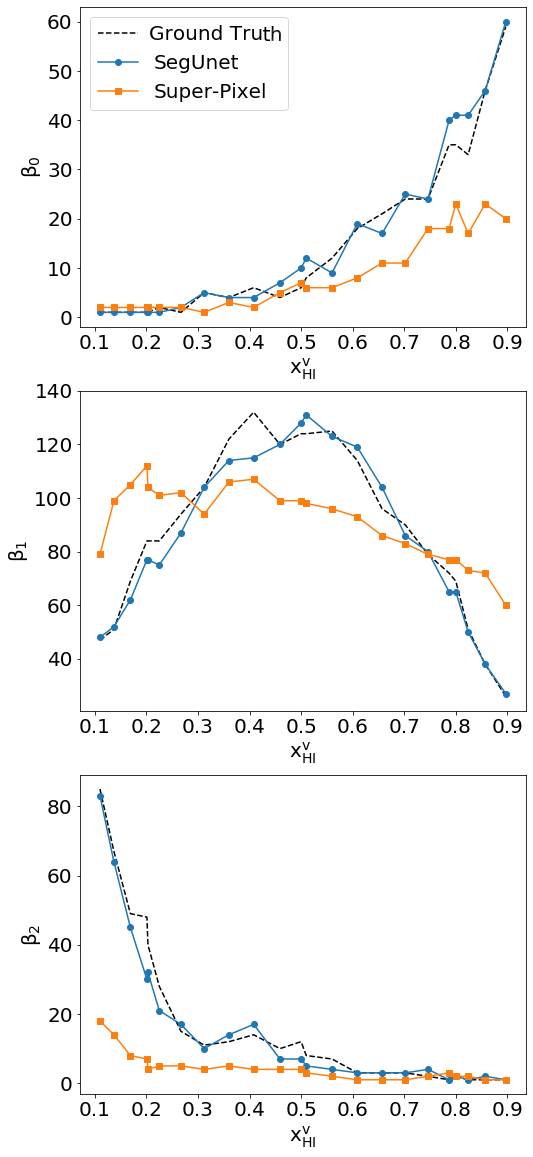}
	\caption{Comparison of the topology of the identified regions with Betti numbers estimated from the original neutral field (black dashed line), the \texttt{SegU-Net} (blue line with circles) and the Super-Pixel method (orange line with squares), for the case of our `fiducial' simulation. The top, middle and bottom panels give $\beta_0$, $\beta_1$ and $\beta_2$ respectively. The Betti numbers recovered with \texttt{SegU-Net} matches the ground truth better than those recovered with the Super-Pixel method.} 
	\label{fig:betti}
\end{figure}

\iffalse
\begin{figure*}
	\centering
	\begin{minipage}{\columnwidth}
	\centering
    	\includegraphics[width=\columnwidth]{Pk_comparison_v}
    	\caption{Dimensionless power spectra of the neutral field from the fiducial simulation as recovered by our network (blue line) and its respective confidence interval (blue shadow). Compared at early, middle and late stage of reionization (from top to bottom $ x^{\rm V}_{\rm HI} = 0.8,\,0.5,\,0.2$) with the same quantity derived from the ground truth (black dashed line) and the Super-Pixel method (orange line). At the bottom of each panel, we show the relative difference compared to the ground truth for both cases, the network and Super-Pixel method.}
    	\label{fig:pk_comparison}
    \end{minipage}
    \hfill
    \begin{minipage}{\columnwidth}
		\centering
    	\includegraphics[width=\columnwidth]{betti_tobs1000}
    	\caption{Comparison of the topology of the identified regions with Betti numbers estimated from the original neutral field (black dashed line), the \texttt{SegU-Net} (blue line with circles) and the Super-Pixel method (orange line with squares), for the case of our `fiducial' simulation. The top, middle and bottom panels give $\beta_0$, $\beta_1$ and $\beta_2$ respectively. The Betti numbers recovered with \texttt{SegU-Net} matches the ground truth better than those recovered with the Super-Pixel method.} 
    	\label{fig:betti}
	\end{minipage}
\end{figure*}
\fi

\subsection{Size Distributions}\label{sec:bsd}
From the 3D tomographic data that will be produced with the upcoming SKA experiment, we will be able to study the size distribution of neutral or ionized region during the EoR. ionized regions are often called bubbles and whereas neutral regions are referred to as islands. The Bubble and Island size distributions (BSDs and ISDs) are useful to derive the properties of reionization and its evolution \citep[][]{Xu2017IslandFAST:Reionization,Giri2019NeutralTomography}. Several approaches were presented to calculate this distribution \citep{Lin2016,Kakiichi2017,Giri2018BubbleTomography}. In this work, we employ the Mean-Free-Path method \citep[MFP; ][]{Mesinger2007EfficientReionization,Giri2018BubbleTomography} to calculate the size distribution ($R \frac{\mathrm{d}N}{\mathrm{d}R}$) of recovered neutral (ISD) and ionized field (BSD). Previous works have demonstrated that this method should be preferred since the calculated size distributions are almost unbiased \citep{Lin2016, Giri2018BubbleTomography}.

In the left and right columns of \autoref{fig:bs_comparison} we show the ISDs, respectively BSDs of the binary fields recovered with \texttt{SegU-Net} (blue line) and Super-Pixel method (orange line) compared to the ground truth (black dashed line). The Super-Pixel method performs best when the simulation is halfway through the reionization process $x^{\rm v}_{\rm HI}=0.5$ (central panel). However, it is considerably less accurate compared to \texttt{SegU-Net}. We show the relative difference with the ground truth in the plots below the ISDs and BSDs. The blue shaded region shows the error on each of the size distributions determined by \texttt{SegU-Net}. In both the ISD and BSD case, the main difference between the two recovered distribution occurs at the earlier $ x^{\rm v}_{\rm HI}=0.8$ (top) and later $ x^{\rm v}_{\rm HI}=0.2$ (bottom) stages of reionization. \texttt{SegU-Net} shows a relative difference of a few per cent while the distributions determined from the Super-Pixel segmentations show relative differences up to 10 per cent for large sizes. 

\begin{figure*}
	\centering
	\includegraphics[width=\textwidth]{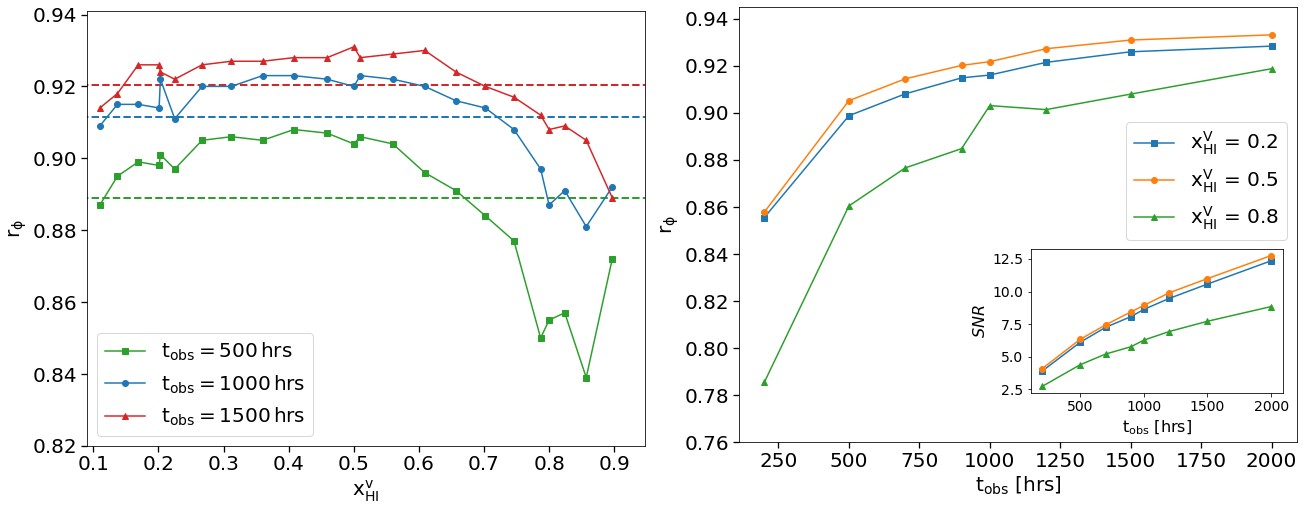}\vskip-2mm
	\caption{\textit{Left panel}: the Matthews correlation coefficient $ r_{\phi}$ of the recovered binary field against its volume-averaged neutral fraction. We compare the prediction set for a high noise level ($ t_{\rm obs}=500\,h$, green line with squares) and low noise level ($ t_{\rm obs}=1500\,h$, red line with triangles) against the noise level employed during the training ($ t{\rm obs}=1000\,h$, blue line with squares). Horizontal dashed lines of the respective colour represent the MCC average score of the reference simulation. \textit{Right panel}:, the evolution of the MCC score for increasing observation time for a set of mock observations with volume-averaged neutral fraction of $ x^{\rm V}_{\rm HI} = 0.2\,(z=7.310),\,0.5\,(z=8.032)$ and $ 0.8\,(z=8.720)$, respectively in blue, orange and green color. In the same panel, an inset plot shows the signal-to-noise ratio (SNR $=\sigma_{\rm 21}/\sigma_{\rm noise}$) of 21-cm images at a resolution corresponding to a maximum baseline of 2 km we achieve for different observation times.}
	\label{fig:phi_coef_tobs}
\end{figure*}

\subsection{Dimensionless Power Spectra}\label{sec:ds21cm}
The dimensionless power spectrum of the neutral field is defined as $\Delta_{\rm x\,x}^2 = k^3 P_{\rm x\,x}(k) / 2 \pi^2$, where $ P_{\rm x\,x}$ is the auto-power spectrum that quantifies the fluctuations due to the distribution of neutral regions. These fluctuations contribute to the 21-cm power spectrum that is observed with radio interferometric telescopes. See for example \citet{Furlanetto2006CosmologyUniverse} and \citet{Lidz2007HigherSpectrum} for descriptions of the fluctuations of the 21-cm signal. In this section, we study the $\Delta_{\rm x\,x}^2$ estimated from the neutral fields recovered from various methods.

In \autoref{fig:pk_comparison}, we consider the `fiducial' simulation at three stages of reionization, which are $x^{\rm V}_{\rm HI}=0.8$ (top panel), $ x^{\rm V}_{\rm HI}=0.5$ (central panel) and $ x^{\rm V}_{\rm HI}=0.2$ (bottom panel). At the mid-point of reionization (central panel), the Super-Pixel method performs well at large scales $k<0.2\,\mathrm{\mpc}^{-1}$ with a relative difference within 25 per cent for lower k-values. The $\Delta_{\rm x\,x}^2$ of the neutral field recovered by the Super-Pixel method at early and late times have the correct shape but differ in magnitude. The $\Delta_{\rm x\,x}^2$ of the neutral field recovered by \texttt{SegU-Net} match the ground truth well at all three stages of reionization. The network maintains a maximum difference compared with the ground truth, of a few tens of per cent at all scales. For $k \lesssim 0.5\,\mathrm{\mpc}^{-1}$, the network uncertainty interval grows to 25-50 per cent relative difference.

\subsection{Betti numbers}\label{sec:betti_numbers}
During reionization ionized bubbles form, grow and connect with each other to form a complex topology \citep[][]{Furlanetto2016ReionizationTheory}. Various studies have proposed topological descriptors for this distribution, such as Euler characteristics \citep[e.g.][]{Friedrich2011} and Betti numbers \citep{Elbers2019topology, Giri2020Betti, Kapahtia2021}. \citet{Giri2020Betti} pointed out that Betti numbers contain more information compared to the Euler characteristics. Therefore in this section we study the zeroth $ \beta_0$, first $ \beta_1$ and second $ \beta_2$ Betti number \citep{Betti1870SopraDimensioni} of the  binary 3D maps recovered by the two feature identification methods. 

$\beta_0$, $\beta_1$ and $\beta_2$ describe the number of isolated ionizing regions, tunnels and isolated neutral regions, respectively. In the top, middle and bottom panels of \autoref{fig:betti}, we show the $\beta_0$, $\beta_1$ and $\beta_2$ values estimated from the recovered binary fields of our `fiducial' model at $x^\mathrm{V}_\mathrm{HI}$ between 0.1 and 0.9. The black, blue and orange curves represent the Betti numbers calculated from the ground truth, recovered field with \texttt{SegU-Net} and Super-Pixel method, respectively. In line with the results for the other quantities discussed above, we find that the topology recovered with \texttt{SegU-Net} is much closer to the ground truth than the one recovered by the Super-Pixel method.

\section{Tests on different instrumental noise levels}\label{chap:other_noise}
We have trained and tested \texttt{SegU-Net} for one specific noise level, corresponding to the theoretically expected noise for $t_\mathrm{\rm obs}=1000$~h with the current design of SKA-Low. However, in practice, the noise level may differ from this, either because the observing time or telescope design is different from our assumptions or simply because the theoretical noise level is not achieved due to complications with the calibration. Therefore, it is important to test to which extent the performance of our network is sensitive to the noise level in the actual data. To change the noise level we choose different observing times, one shorter ($t_\mathrm{\rm obs}=500$ h) and one longer ($t_\mathrm{\rm obs}=1500$ h). The former case corresponds to a noise level $\sqrt{2}$ higher than used in the training set and the latter to a noise level which is $\sqrt{2/3}$ lower.      

In the left panel of \autoref{fig:phi_coef_tobs}, we show the $ r_{\phi}$ coefficient of the recovered binary field against the volume-averaged neutral fraction $x^{\rm V}_{\rm HI}$. We compare the prediction on the reference simulation for the higher ($t_{\rm obs}=500$ h, green line with squares) and lower noise case ($t_{\rm obs}=1500$ h, red line with triangles) with the one using the noise level employed during the training and validation process ($ t_{\rm obs}=1000$ h, blue line with circles). It is evident from the plot that although the noise level does impact the accuracy of the results, we still achieve approximately the same level of precision as in our test case, as commented in \secref{sec:mcc}. In fact, the overall average accuracy, indicated with horizontal dashed lines in \autoref{fig:phi_coef_tobs}, on the simulation of reference is 89 per cent for the higher noise case (green dashed line) and slightly better, 92 per cent, for the lower noise case (red dashed line). In both cases, there is a drop in performance down to 88 per cent accuracy during the early stages of reionization $ x^{\rm V}_{\rm HI}>0.7$, due to the redshift dependency of the simulated noise.

We also want to test how far we can push our \texttt{SegU-Net} trained on data with $ t_{\rm obs}=1000$ h instrumental noise to identify structures in the presence of a higher or lower noise level. In the right panel of \autoref{fig:phi_coef_tobs}, we plot the $ r_{\phi}$ coefficient at different observation times $ t_{\rm obs}$, for three different stages of reionization in our reference simulation, namely for volume-averaged neutral fractions $ x^{\rm V}_{\rm HI}$ is 0.2 (blue line with squares), 0.5 (orange line with circles) and 0.8 (green line with triangles), corresponding to redshift $z\rm=7.310,\,8.032$ and $8.720$. This plot shows that our network performs well for $ t_{\rm obs}\gtrsim 500$ h, where $r_{\phi}\gtrsim 0.85$. The spike in the curve for $ x^{\rm V}_{\rm HI}=0.8$ at $ t_{\rm obs}=1000$~h is due to the fact that this is the noise level for which the network was trained. 

To put our noise level into perspective, the inset plot in the right panel of \autoref{fig:phi_coef_tobs} shows the signal-to-noise ratio (SNR) achieved for different observation times. The SNR is defined as $\sigma_{\rm 21}/\sigma_{\rm noise}$ \citep[e.g.][]{Kakiichi2017}, where $\sigma_{\rm 21}$ and $\sigma_{\rm noise}$ are the standard deviations of the 21-cm signal and noise respectively at the resolution corresponding to a maximum baseline of 2 km. From this we conclude that a good performance, with the same accuracy as the `random' testing set ($r_{\phi}\gtrsim 0.85$), requires a SNR$\gtrsim 3$.

\section{Tests on a fully numerical simulation}\label{chap:c2ray}

We applied our network to mock $\delta T_b$ cubes calculated with the \ctworay code, presented in \secref{sec:c2ray_method}, with a spatial resolution close to the \texttt{21cmFAST} simulations employed in the training process, $\rm2\,Mpc$, in order to obtain the same level of noise per pixel. A visual comparison of the recovered binary field, similar to the results in \secref{sec:visualcomparison}, is shown in \autoref{fig:c2ray_slice}. In the left panel, the red/blue colour indicates the network prediction

We go through the same process presented in \secref{chap:results}. The $ r_{\phi}$ score with \texttt{SegU-Net} is represented by the violet dots with error-bars on the right panel of \autoref{fig:phi_coef}, from left to right we have redshift $z=7.96,\,8.06,\,8.17,\,8.28,\,8.40,\,8.52$ and 8.64 corresponding to a universe with volume-averaged neutral fraction of $ x^{\rm V}_{\rm HI} = 0.17,\,0.29,\,0.42,\,0.57,\,0.70,\,0.81$ and 0.90, green squares represent the score obtained with the Super-Pixel method. As we can see, our neural network is performing with similar accuracy as for the prediction set of semi-numerical simulations as discussed in \secref{sec:visualcomparison}. For $ x^{\rm V}_{\rm HI} \approx 0.55$ \texttt{SegU-Net} performs slightly better than the Super-Pixel method. The Super-Pixel method shows a drop in accuracy at the late ($x^{\rm V}_{\rm HI} < 0.5$) and early ($x^{\rm V}_{\rm HI} > 0.8$) stages of reionization. We do the same comparison with the recovered volume-averaged neutral fraction $ x^{\rm V}_{\rm HI}$, in \autoref{fig:corr} right panel, the green error-bar points are the same data as mentioned above. As we can see, also for the \ctworay simulation, \texttt{SegU-Net} recovered quantity resides within the 0.5-$\sigma$ confidence interval (violet dots with error-bars). For the Super-Pixel results, this is true only for $x^{\rm V}_{\rm HI}= 0.57,\,0.70$ and 0.81, with approximately the same precision as \texttt{SegU-Net} in the case of $x^{\rm V}_{\rm HI}=0.57$ and slightly better results for $x^{\rm V}_{\rm HI}= 0.70$.

\section{Discussion \& conclusions}\label{chap:discussion}
This work has developed a convolutional neural network based on the U-Net architecture, which can be used to segment redshifted 21-cm image observations into neutral and ionized regions. We have shown that this application of deep learning provides a fast and stable method that significantly improves the identification of ionized/neutral regions during the epoch of reionization over previously proposed methods. To train our network, we employ a large set of simulated mock observations of the 21-cm signal. 

Our image segmentation network, \texttt{SegU-Net}, also contains an uncertainty estimator. This uncertainty estimator is a simple but efficient application of the test-time augmented (TTA) technique. With this uncertainty estimator, our network can create a pixel by pixel error map during the prediction process. The pixel by pixel error map can later be used to determine the error in any quantity derived from the segmentation. 

Once the network has been trained, the binary field's extraction is swift. In our case for simulations of volume $(256\,\rm\mpc)^3$ and mesh-grid of $128^3$, a run in serial on a \texttt{Intel\textregistered~Core\texttrademark i7-6500U CPU @ 2.5 GHz} processor and using a 16 Gigabytes of \texttt{RAM} takes between 5 to 10 seconds. Including the pixel-error map calculation increases the computing time to approximately 10 minutes. For comparison, the Super-Pixel method typically requires several minutes to extract the binary field, where the actual time depends on the image pixel resolution and the number of Super-Pixels employed. The computational speed of our network thus makes it particularly useful as a component in a Bayesian statistical inference framework to perform EoR parameter estimation using tomographic statistics \citep[e.g.][]{Gazagnes2021Inferring}.

We compare the accuracy of our approach with a feature finding method from the field of image processing, the so-called Super-Pixel method, which \cite{Giri2018BubbleTomography} showed to be superior over simple thresholding methods. The results show that our neural network can identify neutral regions in the mock observations at least as well and often much better than the Super-Pixel method. We show a visual comparison and the resulting pixel per pixel error map tested on our `fiducial' model. This error map gives valuable insight into the parts of the image that are hard to recover and helps us check for over-fitting.

We studied the accuracy of a range of derived quantities from the recovered binary fields, comparing the performance of \texttt{SegU-Net} with the Super-Pixel method. These quantities are the volume-averaged ionization fraction --- the evolution of which provides the reionization history, the size distribution of the ionized (BSD) and neutral (ISD) regions, the dimensionless power spectra of the recovered binary fields and the three Betti numbers, which quantify the topology of the segmented data sets. For all quantities, we find that the \texttt{SegU-Net} results are more accurate than the Super-Pixel results, especially for the early and late stages of reionization, where the Super-Pixel method often struggles to produce accurate results.   

Machine learning methods generally are sensitive to the properties of the training set. Therefore, we tested \texttt{SegU-Net} on input data with different properties than the training set. First, we analysed the performance on data sets with different noise levels than the network was trained. We found that \texttt{SegU-Net} yields accurate results for data sets in which the noise level is characterised by an observing time of $ t_{\rm obs}>500\,h$, which approximately corresponds to an SNR $\gtrsim 3$. Second, we applied the network to mock observations calculated from the results of a fully numerical reionization simulation, rather than the semi-numerical simulations used to train the network. We find that \texttt{SegU-Net} achieves the same level of accuracy when applied to this data set and therefore is not sensitive to the type of simulation employed during the training process. 

We want to point out that similar efforts are being made by \cite{Gagnonhartman2021recovering}. They focus on reconstructing the segmented maps of ionized and neutral regions in the context of foreground mitigation using the foreground avoidance method \citep[e.g.][]{2011PhRvD..83j3006L,2014ApJ...782...66P}, and also consider the possibility of doing so with instruments that are not optimised for imaging such as HERA. We include the effect of instrumental noise and study in great detail the summary statistics of the reconstructed binary maps and the dependency of the results on the noise level. In the future, we will extend our study to include the impact of foreground mitigation strategies while recovering the summary statistics.

Here we assumed the spin temperature to be saturated ($T_\mathrm{S}\gg T_\mathrm{CMB}$). However, it is possible to have such a scenario where this assumption fails, especially during the time when reionization starts. In the future, we will evolve our \texttt{SegU-Net} to identify \HII regions in such scenarios. Even though our network is built to identify \HII regions, U-Net architecture can be trained to identify any interesting feature. Before reionization started, the luminous sources heated the IGM and left its impact on the 21-cm signal \citep[e.g.][]{Ross2017SimulatingDawn,Ross2019EvaluatingDawn}. The U-Net architecture can also be trained to identify these heated regions.

\section*{Acknowledgements}
The authors would like to thank Leon Koopmans for useful discussions and comments. We also acknowledge helpful discussion with Adrian Liu and collaborators. MB is supported by PhD Studentship from the Science and Technology Facilities Council (STFC) and appreciates the Oskar Klein Center at Stockholm University for hospitality during the completion of this work. This work was possible thanks to the STFC Long Term Attachment (LTA) travel grant. ITI is supported by the Science and Technology Facilities Council (grant numbers ST/I000976/1 and ST/T000473/1) and the Southeast Physics Network (SEPNet). GM is supported in part by the Swedish Research Council grant 2020-04691. We acknowledge PRACE for awarding us access to the KAY facility hosted by the Irish Centre for High-end Computing (ICHEC) and the GALILEO hosted by the Super Computer Application and Innovation (SCAI) in collaboration with the CINECA consortium. The authors gratefully acknowledge the Gauss Centre for Supercomputing e.V. (www.gauss-centre.eu) for partly funding this project by providing computing time through the John von Neumann Institute for Computing (NIC) on the GCSSupercomputer JUWELS at Juelich Supercomputing Centre (JSC). The deep learning implementation was possible thanks to the application programming interface of \texttt{Tensorflow} \citep{Tensorflow2015} and \texttt{Keras} \citep{Chollet2017}. The algorithms and image processing tools operated on our data were performed with the help of \texttt{NumPy} \citep{numpy2020}, \texttt{SciPy} \citep{scipy2020} and \texttt{scikit-image} packages \citep{scikit2014}. All figures were created with \texttt{mathplotlib} \citep{Hunter2007}.

\section*{Data Availability} 
The data underlying this article is available upon request, and can also be re-generated from scratch using the publicly available \texttt{21cmFAST}, \texttt{CUBEP$^3$M}, \ctworay and \texttt{Tools21cm} code. The \texttt{SegU-Net} code and its trained network weights are available on the author's \texttt{GitHub} page: \url{https://github.com/micbia/SegU-Net}.
%%%%%%%%%%%%%%%%%%%% REFERENCES %%%%%%%%%%%%%%%%%%
\bibliographystyle{mnras}
\bibliography{segnet_ref,mendeley_sg}
%%%%%%%%%%%%%%%%% APPENDICES %%%%%%%%%%%%%%%%%%%%%
%\vspace*{\fill}
%\pagebreak
\appendix
\section{\texttt{SegU-Net} Hidden Layer Outputs} \label{app:hidden_layer}
\begin{figure}
    \centering
	\includegraphics[width=\columnwidth]{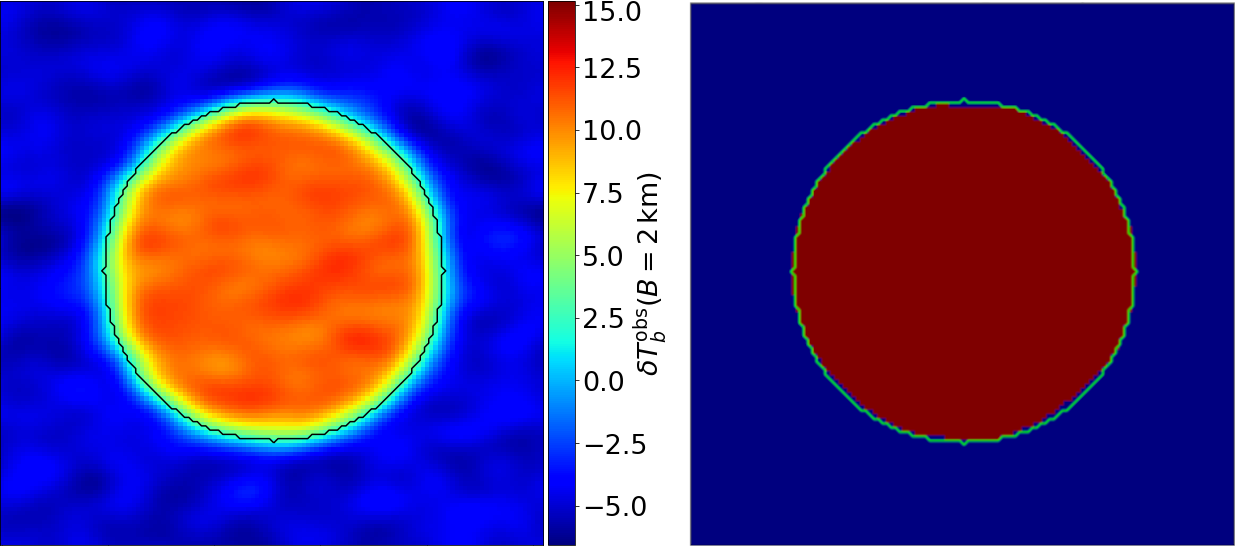}
	\caption{Test \texttt{SegU-Net} on a spherical ionized region. \textit{Left panel}: slice through the input image. The colour map shows the differential brightness temperature, and the black contour shows the boundary between neutral and ionized regions. \textit{Right panel}: the recovered binary field with \texttt{SegU-Net}. The green contour represents the same boundary again. The identified neutral and ionized regions are indicated in red and blue, respectively.}
	\label{fig:sphere}\vskip1cm

	\includegraphics[width=\columnwidth]{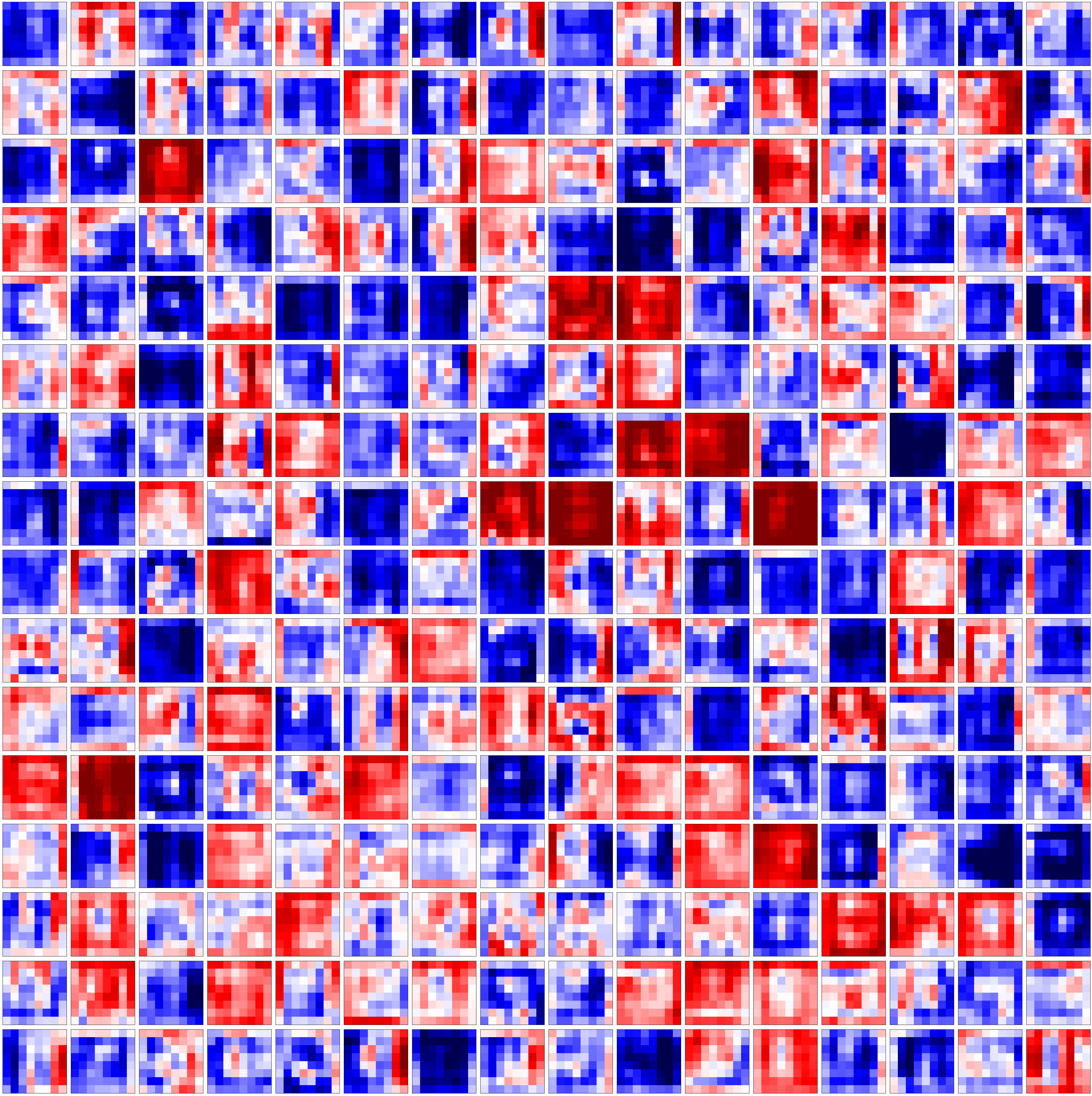}
	\caption{Visual representation of \texttt{SegU-Net}'s low dimensional latent space (bottom layer), which contains information about the extracted features of our test input image.}
	\label{fig:layer36}
\end{figure}

We test \texttt{SegU-Net} to see if it can recover the binary field for a simple case, namely a single spherical neutral region. We assume a uniform density field at $z=8.032$ and calculate the differential brightness temperature with \autoref{eq:dTb}, adding noise corresponding to $t_{\rm obs}=\rm1000\,h$ and reducing the resolution to correspond to a maximum baseline of $B \rm=2\,km$. In \autoref{fig:sphere}, we show the input image of the sphere (left panel) and the corresponding recovered binary field by \texttt{SegU-Net} (right panel). The black contour in the left panel and the green contour in the right panel show the true boundary of the sphere. For this test \texttt{SegU-Net} achieves an accuracy of 98 per cent.

In \autoref{fig:layer36} we show the output of the bottom hidden layer of \texttt{SegU-Net}, which is the last layer of the left part of the U-shaped in \autoref{fig:architecture}. The colour coding is such that blue correspond to negative, red to positive and white to zero values. This output gives a visual representation of the low dimensional latent space of our network encoder. In our case, this consists of 256 images, where each corresponds to a convolutional kernel and contains information about the image's extracted features. The encoder path contracts the input image from an original mesh size of 128$^2$ down to 8$^2$. The information contained in the latent space is then expanded by \texttt{SegU-Net} decoder into a binary map of the same size as the input image (see right panel of \autoref{fig:sphere}).

\begin{figure*}
    \centering
	\includegraphics[width=\textwidth]{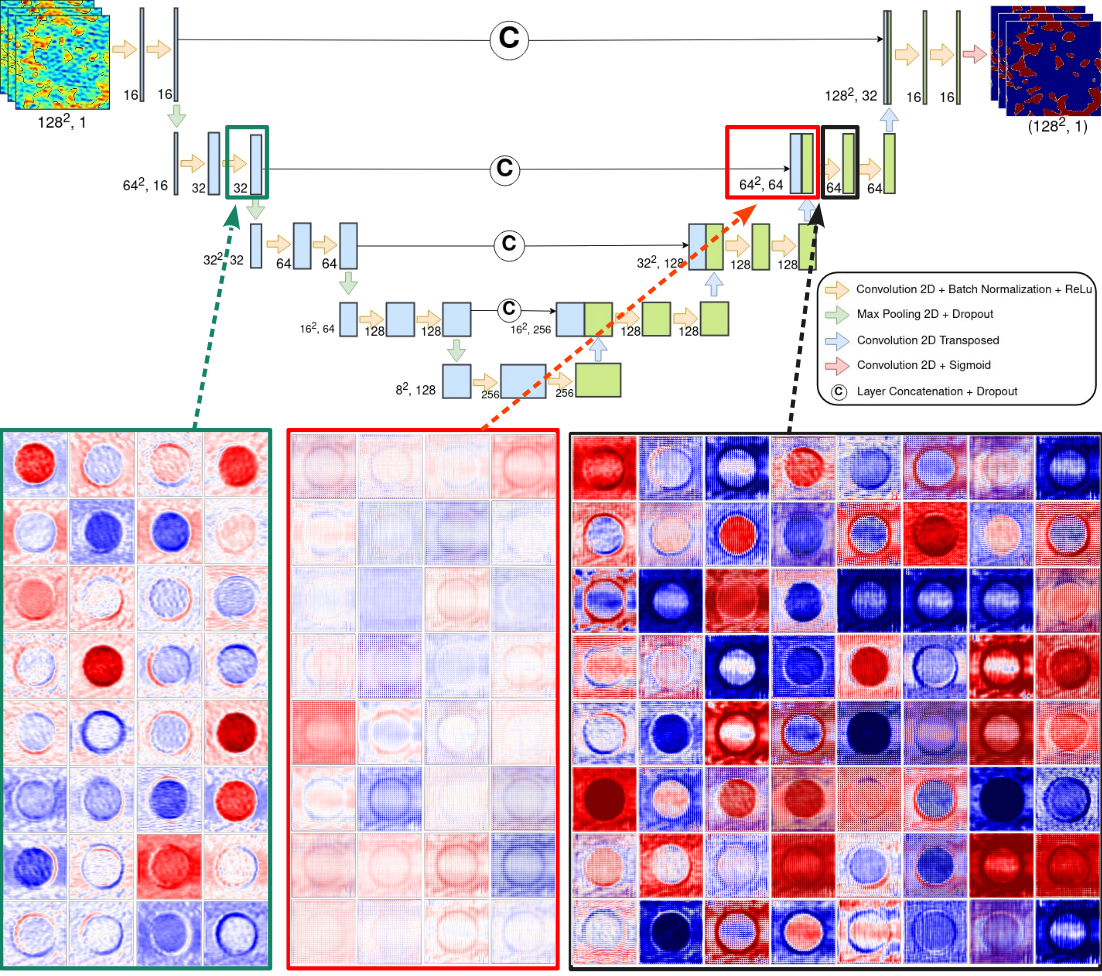}\vskip-2mm
	\caption{Example of skip connection between encoder and decoder levels. The top panel shows the architecture of our network. The bottom panels display the output of three hidden layers. On the left (green dashed line), a convolutional block (\texttt{ConvBlock}) output is interconnected with the output of the second to last up-sampling operation (central panel, red dashed line). The right-most panel shows the result of the merge after a convolution block (black dashed line).}
	\label{fig:example_layer}
\end{figure*}

\section{Skip Connection between encoder and decoder levels} \label{app:concatenation_example}
The main advantage of a U-shaped network is that it overcomes the bottleneck limitation encountered by auto-encoder networks (a classical encoder/decoder architecture) by adding interconnections between the descending (encoder) and ascending (decoder) paths \citep{Long2014, Ronneberger2015}. These connections allow feature representations to pass through the bottleneck (bottom layer) and avoid loss of information due to contraction.

In \autoref{fig:example_layer}, we show a visual example of interconnections between the encoder (left part of the U-shape) and the decoder (right part). The top panel shows a schematic representation of our network architecture, and the bottom part displays a visual output of three hidden layers for the test case of a sphere. The left-most panel (connected by a green dashed line) shows the output of the second convolutional block in the encoder's second level. This block consists of 32 kernels with a mesh size of 64$^2$. At this level, the shape and form of the input image are still visible.
The centre panel (connected by a red dashed line) shows the result of the second to last up-sampling operation of the decoder. The number of kernels and mesh size match with the corresponding encoder layers. The skip connection merges the encoder and decoder-level output for a total of 64 images with mesh 64$^2$. The right-most panel (connected with a black dashed line) shows the concatenation after a convolutional block. The effect of the up-sampling operation is still visible.

%%%%%%%%%%%%%%%%%%%%%%%%%%%%%%%%%%%%%%%%%%%%%%%%%%
%\bsp	% typesetting comment
\label{lastpage}
\end{document}